\documentclass[%
 aip,
 amsmath,amssymb,
preprint,%
]{revtex4-1}

\usepackage{bm}
\usepackage{color}
\usepackage{dcolumn}
\usepackage{etoolbox}
\usepackage{float}
\usepackage[T1]{fontenc}
\usepackage{graphicx}
\usepackage[utf8]{inputenc}
\usepackage{mathptmx}
\usepackage{multirow}
\usepackage{stmaryrd}  
\usepackage[caption = false]{subfig}

\makeatletter
\def\@email#1#2{%
 \endgroup
 \patchcmd{\titleblock@produce}
  {\frontmatter@RRAPformat}
  {\frontmatter@RRAPformat{\produce@RRAP{*#1\href{mailto:#2}{#2}}}\frontmatter@RRAPformat}
  {}{}
}%
\makeatother
\begin{document}

\preprint{AIP/123-QED}

\title[Drag prediction using effective models]{Prediction of drag components on rough surfaces using effective models}
\author{Sahaj Jain}%
\author{Y. Sudhakar}
 \email{sudhakar@iitgoa.ac.in}
\affiliation{ 
School of Mechanical Sciences, Indian Institute of Technology Goa, Farmagudi, Goa -- 403 401,  India.
}%

\date{\today}

\begin{abstract}
Owing to the multiscale nature and the consequent high computational cost associated with simulations of flows over rough surfaces, effective models are being developed as a practical means of dealing with such flows. Existing effective models focus primarily on accurately predicting interface velocities using the slip length. Moreover, they are concerned mainly with flat interfaces and do not directly address the drag computation. In this work, we formulate the Transpiration-Resistance model in polar coordinates and address the challenge of computing drag components on rough surfaces. Like the slip length, we introduce two constitutive parameters called shear and pressure correction factors that encompass information about how the total drag is partitioned into viscous and pressure components. Computation of these non-empirical parameters does not necessitate solving additional microscale problems; they can be obtained from the same microscale problem used for the slip-length calculation. We demonstrate the effectiveness of the proposed parameters for the Couette flow over rough surfaces. Moreover, using the flow over a rough cylinder as an example, we present the accuracy of predicting interface velocity and drag components by comparing the effective model results with those obtained from geometry-resolved simulations. Numerical simulations presented in this paper prove that we can accurately capture both viscous and pressure drag over rough surfaces for flat- and circular-interface problems using the proposed constitutive parameters. 
\end{abstract}

\maketitle

\section{Introduction}
\label{sec:intro}
Transport phenomena occurring over rough and patterned surfaces are crucial in various practical applications.
Notable examples include riblets on shark skins~\cite{bechert1989}, superhydrophobic surfaces~\cite{truesdell2006}, contact line modelling~\cite{smith2018}, flows in microfluidic devices~\cite{stone2004} and biomechanics~\cite{owen2020}.

The characteristic feature of these applications is that the length scale of the roughness elements ($l$) is several orders of magnitude smaller than the macroscopic length scale ($H$) of the surrounding fluid flow. Due to a very fine computational mesh requirement near the boundary, this intrinsic multiscale nature makes the geometry-resolved numerical simulations (denoted hereafter as DNS) of such configurations an impossible task. The practical difficulty of performing DNS has motivated the development of effective or homogenized models that replace the rough surface with a fictitious smooth wall.  In effective models, topographical details of the roughness elements are discarded, which significantly reduces the computational cost. Such models aim to describe the macroscopic behaviour of fluid flows by capturing microscale-averaged flowfields. 

In effective models, the physical influence of surface roughness is represented with appropriate boundary conditions at the fictitious smooth wall.  Due to their practical importance, the task of deriving accurate effective boundary conditions has received great attention in the literature.  The most commonly used boundary condition in the effective model is the Navier-slip condition~\cite{sarkar1996,bolanos2017}
\begin{equation}
u=\lambda\frac{\partial u}{\partial y},
\label{eqn:intronavslip}
\end{equation}
where $\lambda$ is called the slip length; it is the distance into the wall where the linearly extrapolated velocity goes to zero. The shear-induced slip velocity condition, in addition to rough surfaces, is also used in superhydrophobic surfaces~\cite{min2004} and liquid-infused surfaces~\cite{kim2020}.  Similar boundary conditions are also used to model free-fluid interacting with porous media~\cite{beavers1967,lacis2016}.  Research papers addressing flow over riblets denote the slip length as the protrusion height~\cite{bechert1989}.  A second-order boundary condition, which contains nonlinear terms for slip velocity, is also proposed\cite{achdou1998}.

The condition expressed in equation~\eqref{eqn:intronavslip} works well for isotropic rough surfaces, with the scalar parameter $\lambda$ representing all the effects of roughness elements. For three-dimensional flow over anisotropic surfaces, the tangential flow need not coincide with the mean shear~\cite{stroock2002,bazant2008}. Formulations for such surfaces are dealt with by deriving a  tensorial version of the Navier-slip boundary condition and introducing the associated mobility tensor~\cite{kamrin2010}.  Symmetry properties of the mobility tensor are proved~\cite{kamrin2011}. A comprehensive overview of the subject, covering effective models of porous and poroelastic surfaces, is given by Bottaro~\cite{bottaro2019}. 

Effective models discussed above represent the effect of surface roughness with the Navier-slip condition for tangential velocity and a zero transpiration velocity. Based on the fundamental principle of mass-conservation, Ugis et al.~\cite{lacis2020} have shown that the transpiration is zero only when the tangential velocity (or in other words, shear at the effective wall) is constant along the entire length of the interface.  
They proposed the transpiration-resistance~(TR) model that enables accurate computation of transpiration and tangential velocity components over rough/porous surfaces.
Although the magnitude of transpiration velocity is much smaller than the tangential component, it has been demonstrated on a channel flow that the transpiration is crucial to model the physics of turbulent flows.  The mathematical derivation of the TR model, based on higher-order multiscale homogenization,  is provided by Sudhakar et al. \cite{sudhakar2021}.  The applicability of the TR model in turbulent regimes is investigated~\cite{khorasani2021}, and it is shown to reproduce near-wall turbulent modifications in the transitionally rough regime. In a recent work, Bottaro \& Naqvi\cite{bottaro2020} obtained higher-order interface conditions that are correct upto $\mathcal{O}(\epsilon^3)$.

Available works on effective models mainly address configurations in which  the fictitious interface is aligned with one of the axes of Cartesian coordinates.  In a recent work\cite{zampogna2019}, a generalized slip condition over rough surfaces is derived.  This work considers flow over a sphere coated with hexagonal lattice of cylindrical roughness elements, which is a fully three-dimensional test case.  
Results from the effective model are compared with DNS in laminar and turbulent flow regime.  Based on an effective model to simulate interaction between an incompressible fluid and thin permeable membranes\cite{zampogna2020}, a recent study\cite{ledda2021} reported homogenization-based design and optimisation of membranes consisting of permeable cylindrical shells. To the best of our knowledge,  Zampogna et al.\cite{zampogna2019} is the only reported work that comprehensively analysed the accuracy of effective models for rough surfaces to non-flat interfaces.  It is observed that the addition of roughness increases the pressure drag significantly  and reduces the viscous component. While the effective model is shown to predict the total drag with reasonable accuracy, it is unable to represent components of drag.  This is remarked as one of the limitations of  effective models.   The present paper addresses this limitation by introducing relevant constitutive parameters within the TR model’s framework. 

\begin{figure*}
\subfloat[DNS]{\includegraphics[scale=0.4]{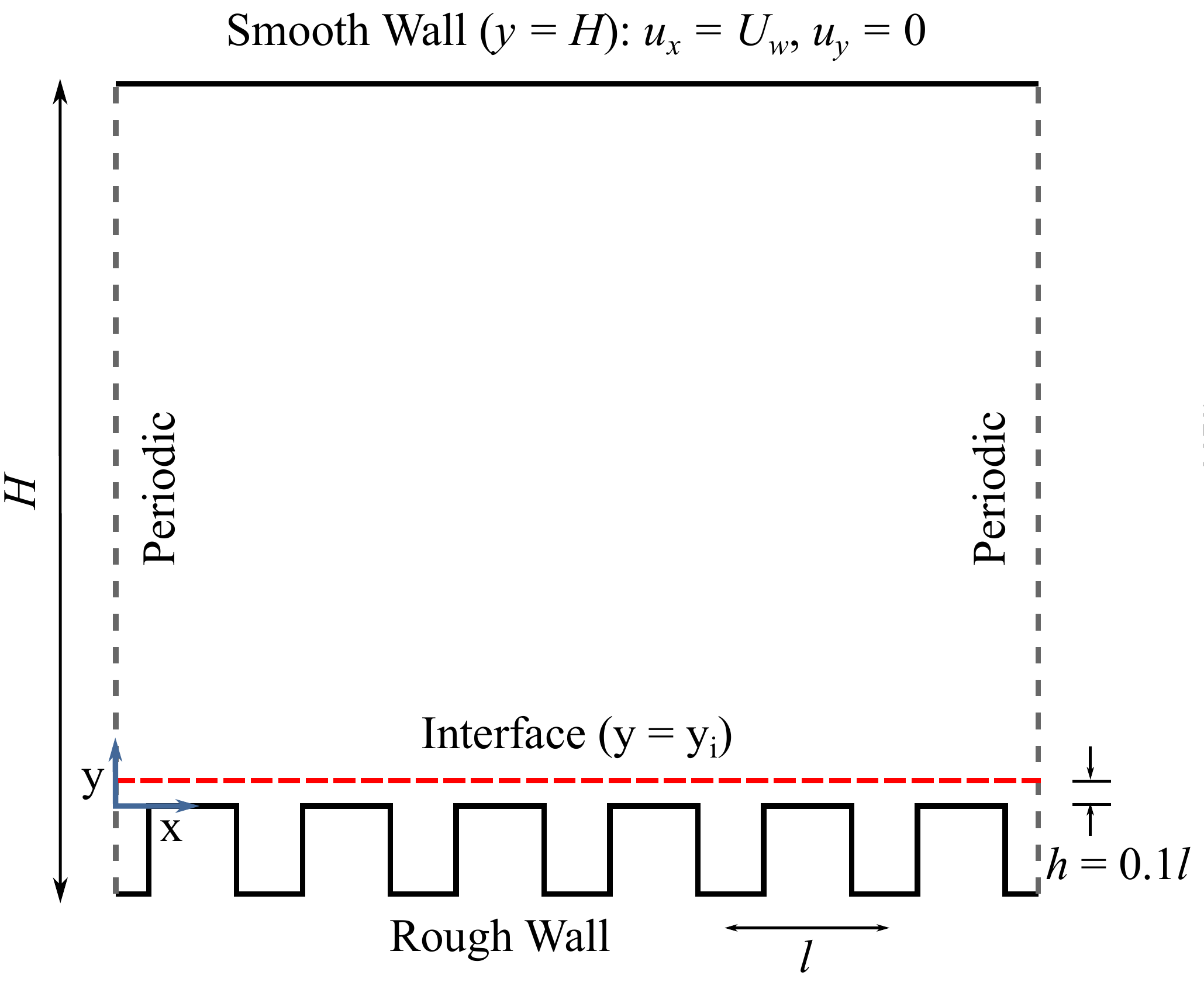}} \hspace{0.5cm}
\subfloat[Effective Model]{\includegraphics[clip, scale=0.355]{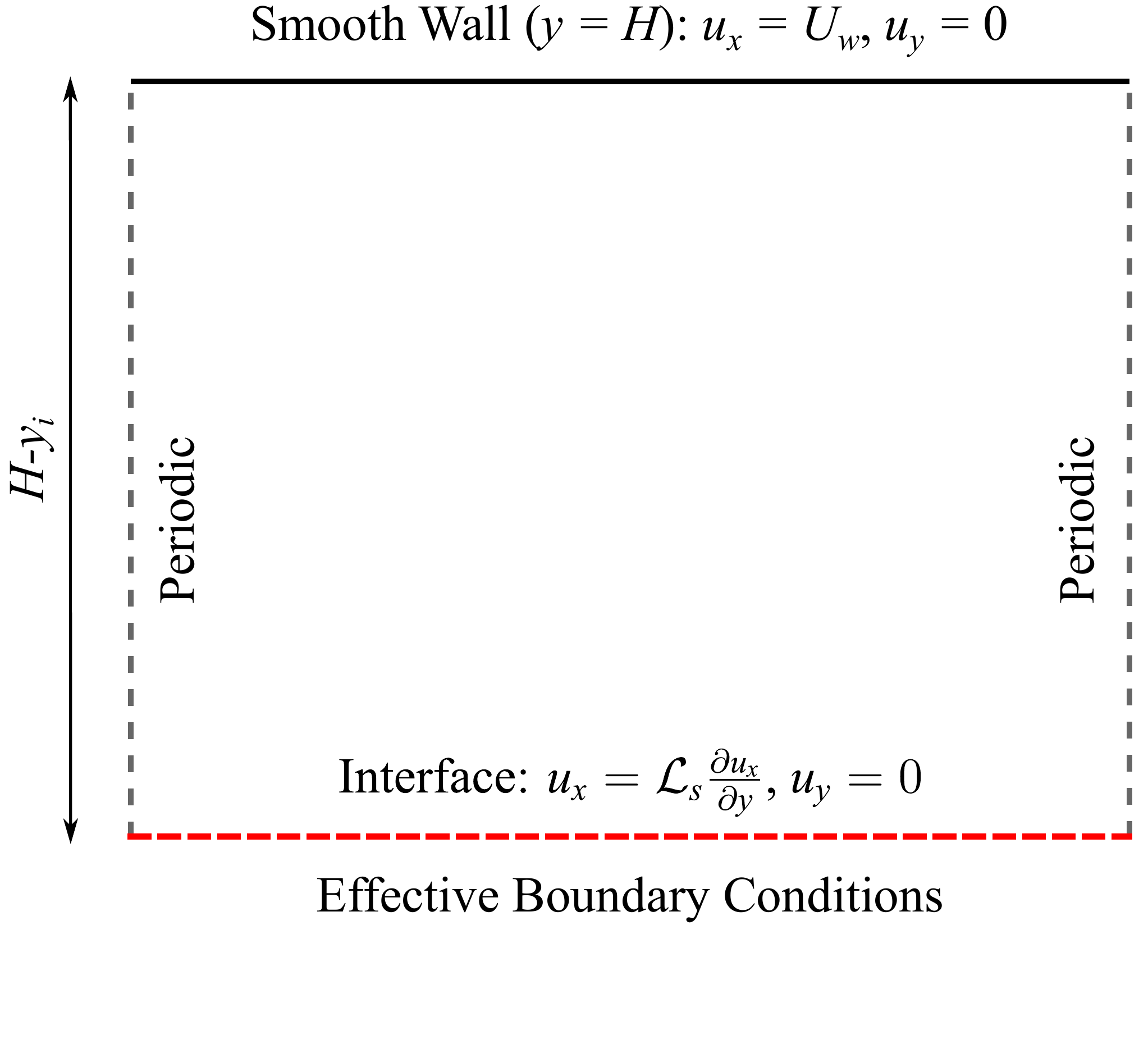}}
\caption{Geometry and boundary conditions for the Couette flow (a)~DNS (b)~Effective model.}
\label{fig:couetgeom}
\end{figure*}

Surface roughness features have an enormous impact on drag reduction in the form of riblets, superhydrophobic,  and liquid-infused surfaces.  Despite this importance, the focus of the majority of existing studies is on capturing velocity components accurately on the fictitious interface.  As reported~\cite{zampogna2019}, effective models can predict the total drag but not how the drag is partitioned into pressure- and viscous components. This aspect has not received attention in the literature. The contributions of this paper are two-fold : (1)~we introduce constitutive parameters to accurately capture  viscous and pressure drag on rough surfaces using effective models,  and  (2)~we formulate the TR model~\cite{lacis2020} in polar coordinates and demonstrate the accuracy of the proposed constitutive parameters both for flat and curved interfaces. Moreover, we provide source codes used to obtain results presented in this paper in a public repository\cite{bitbucket}.

Since we work in the framework of the TR model, assumptions made while deriving the model will carry over to this work. They are as follows. (1)~The surface is coated with ordered roughness elements, (2)~There is a clear scale separation, $\eta={l/H}\ll 1$ and (3)~Inertial effects are neglected in the interface region.

This paper is organized as follows.  The constitutive parameters called stress correction factors, relevant for predicting drag components, are introduced in section~\ref{sec:couette}, and they are validated for problems with flat interfaces.  In section~\ref{sec:cyl}, TR model formulations are derived in polar coordinates, and the effectiveness of stress correction factors is demonstrated for flow over rough circular cylinders.  Conclusions are presented in section~\ref{sec:conclusion}.  Effects of neglecting the interface curvature in the effective model is discussed for the rough cylinder problem in appendix~A.
\section{Couette flow over rough surfaces}
\label{sec:couette}
In this section, we consider the well-known Couette flow problem. The stationary bottom wall is covered with rough elements, while the top smooth wall moves with the prescribed velocity~$U_w$~(figure~\ref{fig:couetgeom}(a)). In the effective model, we replace the rough surface with a smooth fictitious interface located at a vertical distance of $y_i$ above the crest plane of rough inclusions~(figure~\ref{fig:couetgeom}(b)). In this example, the interface is aligned with one of the coordinate axes, which is usually the case considered in existing studies. This enables us to use available interface conditions for the effective model. The current section focuses on the computation of drag components on the rough wall using an effective model.
\subsection{Interface velocity formulation}
The effect of roughness elements is represented in the homogenized model by effective boundary conditions. As mentioned earlier, in this work, we make use of the TR model~\cite{lacis2020}, which involves the following conditions at the interface
\begin{subequations}
\begin{align}
    u_x &= \mathcal{L}_{s}\frac{\partial u_x}{\partial y},\label{eqn:slipcouette}\\
    u_y & = -\mathcal{M}\frac{\partial u_x}{\partial x},
\end{align}
\end{subequations}
where $\mathcal{L}_{s}$ and $\mathcal{M}$ are slip length and transpiration length respectively.  These constitutive parameters  contain information about the geometry of roughness elements.  For this particular example, the tangential velocity~($u_x$) remains constant along the interface, and hence the transpiration velocity at the interface is identically zero, $u_y=0$.

Starting from the Navier-Stokes equations, with equation~\eqref{eqn:slipcouette} as the boundary condition at the interface, we can obtain an analytical expression for the velocity profile of the effective model
 \begin{equation}
     u_x=U_w\left(\dfrac{\mathcal{L}_s+y-y_i}{\mathcal{L}_s+H-y_i}\right),
     \label{eq:uanalyt}
 \end{equation}
 and for the constant shear stress across the channel
 \begin{equation}
    \tau _{xy}=\mu\dfrac{d u_x}{d y}=\dfrac{\mu U_w}{\mathcal{L}_s+H-y_i}.
    \label{eq:tauanalyt}
 \end{equation}
 Hence performing simulations using an effective model is not necessary for this particular example, and we use above relations to compute the interface velocity as well as the drag force acting on the fictitious interface in the effective model.
 
\begin{figure}
\subfloat[]{\includegraphics[scale=0.8]{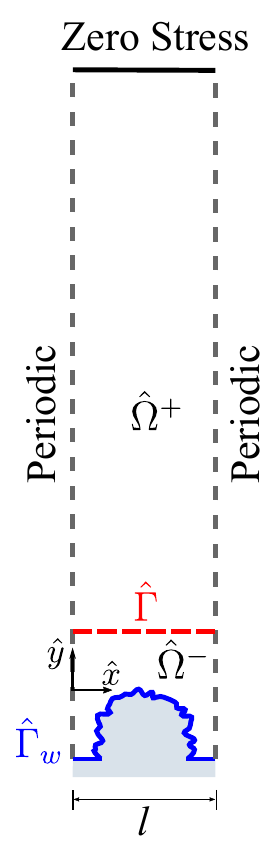}} \hspace{0.5cm}
\subfloat[]{\includegraphics[scale=0.8]{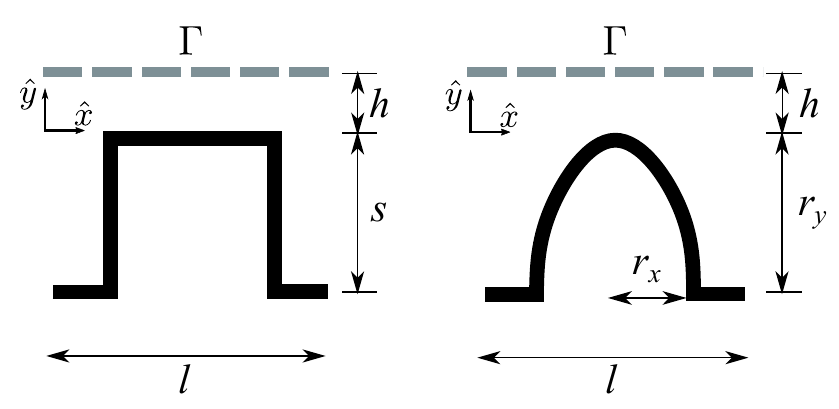}}
\caption{Microscale problem presented in equation~\eqref{eqn:couetLprob}. (a)~Computational domain, and (b)~Geometry of square and ellipse-shaped roughness elements.}
\label{fig:couetmicroprob}
\end{figure}

 In order to use the above formulation, we need to compute the slip length $\mathcal{L}_s$ that depends on the geometrical details of rough elements. The computation of $\mathcal{L}_s$ involves solving a microscale problem on an interface cell~\cite{lacis2016,lacis2020,bottaro2020,sudhakar2021},  as shown in figure~\ref{fig:couetmicroprob}(a). The width of the interface cell is a equal to the microscale~($l$) and its height above the interface in microscale problem~($\hat{\Gamma}$) is equal to 5 times that of $l$. Boundary conditions used in the microscale problems are also given in the figure. The governing equations for the microscale problem~\cite{sudhakar2021} are given below.
 \begin{subequations}
 \begin{align}
     \nabla\cdot L^{\pm}_{ix}&=0 \quad \textrm{on } \hat{\Omega}^{\pm}\\
     -\nabla B^{\pm}_x + \Delta L^{\pm}_{ix} &=0 \quad \textrm{on } \hat{\Omega}^{\pm}\\
     \llbracket L_{ix} \rrbracket &= 0 \quad \textrm{on } \hat{\Gamma}\\
     \left\llbracket -B_x+2\frac{\partial L_{yx}}{\partial y} \right\rrbracket &= 0 \quad \textrm{on } \hat{\Gamma}\\
     \left\llbracket \frac{\partial L_{xx}}{\partial y} + \frac{\partial L_{yx}}{\partial x}\right\rrbracket &= -1 \quad \textrm{on } \hat{\Gamma} \label{eqn:couetLprob_e}
 \end{align}
 \label{eqn:couetLprob}
 \end{subequations}
where $i\in \{x,y\}$. No-slip condition is applied on the roughess elements~($\hat{\Gamma}_w$). In the above equation system and in the upcoming sections, we use the following compact notation
\begin{equation}
     a^{\pm}=c^{\pm}+d\quad \text{ on } \hat{\Omega}^{\pm}\nonumber\\
\end{equation}
 to denote the following
\begin{align}
   a^{+}&=c^{+}+d \quad \text{ on } \hat{\Omega}^+ \nonumber\\
   a^{-}&=c^{-}+d \quad \text{ on } \hat{\Omega}^-. \nonumber
\end{align}
The jump operator is defined as,
\begin{equation}
\llbracket a \rrbracket=a^+ - a^-.
\end{equation}

The solution of the microscale problem is driven by unit shear forcing at the interface~\cite{lacis2016,lacis2020,sudhakar2021}, as given in equation~\eqref{eqn:couetLprob_e}. By numerically solving equation~\eqref{eqn:couetLprob}, we will get the fields of $L_{ix}$ and $B_x$ which are analogous of velocity vector and pressure in a Stokes system. 
 The slip length can be obtained from the fields by performing an averaging at the interface
\begin{equation}
  \mathcal{L}_s=\left\langle L^+_{xx} \right\rangle _s,
  \label{eq:Ls_couet}
\end{equation}
 where \( \left\langle \cdot \right\rangle _s \) is the surface averaging operator defined as
\begin{equation}
\left\langle a\right\rangle _s =\dfrac{1}{\left|\hat{\Gamma}\right|}{\int_{\hat{\Gamma}} a d \hat{\Gamma}}.
\end{equation}
In the above expression, $\left|\hat{\Gamma}\right|$ denotes the length (area in case of 3D) of $\hat{\Gamma}$. Once the slip length is computed, we can get the homogenized solution of the Couette flow problem using equations~\eqref{eq:uanalyt} and \eqref{eq:tauanalyt}.
\subsection{Computation of drag}
 As will be shown below, computing components of drag directly on the fictitious interface will lead to significant errors. In order to get accurate drag components on rough surfaces, we introduce two constitutive parameters associated with rough surfaces in this work, namely the pressure correction factor~($\mathcal{P}_c$) and the shear correction factor~($\mathcal{S}_c$). These two parameters are denoted together as stress correction factors. These factors can be computed from the microscale problem using
 \begin{subequations}
 \begin{align}
     \mathcal{S}_c&=\dfrac{1}{\left|\hat{\Gamma}\right|}\int_{\hat{\Gamma}_w}{\hat{\bm{n}}\cdot\bm{\tau}^{L_{ix}}\cdot\hat{\bm{x}}\, d\hat{\Gamma}_w},\\
    \mathcal{P}_c&=\dfrac{1}{\left|\hat{\Gamma}\right|}\int_{\hat{\Gamma}_w}{-B_x\hat{\bm{n}}\cdot\hat{\bm{x}}\, d\hat{\Gamma}_w},
 \end{align}
 \label{eq:scfactors_couet}
 \end{subequations}
where $\hat{\Gamma}_w$ is the rough wall surface in the microscale problem as can seen in figure~\ref{fig:couetmicroprob}, $\hat{\bm{n}}$ is the normal vector to $\hat{\Gamma}_w$ pointing towards $\hat{\Omega}^-$,  $\hat{\bm{x}}$ is the unit vector along $x-$direction, and $\bm{\tau}^{L_ix}$ is the viscous stress tensor from the microscale problem defined as
 
 \begin{equation}
     \tau_{mn}^{L_ix}=\left(L_{m1,n}+L_{n1,m}\right).
 \end{equation}
 
 The physical meaning of these factors can be understood by considering the microscale problem. Due to the applied unit shear, the rough surface~($\hat{\Gamma}_w$) experiences resistance due to viscosity as well as due to the shape of the elements (pressure drag). By employing a simple force-balance analogy on the interface cell, taking account of the fact that unit shear is applied at the interface, we can show that $\mathcal{P}_c+\mathcal{S}_c=1$. This observation implies that the stress correction factors introduced in the present work extract information about how the total drag on the rough surface is partitioned into viscous and pressure components. 
 
 Components of drag on an effective model can be computed by
 \begin{subequations}
 \begin{align}
     F_v &= \mathcal{S}_c\int_{\Gamma}{\tau_{xy} dS},\\
     F_p & = \mathcal{P}_c\int_{\Gamma}{\tau_{xy} dS},
 \end{align}
 \label{eq:dragcouet}
 \end{subequations}
 where $\Gamma$ is the fictitious interface used in the effective model, and $\tau_{xy}$ can be computed using equation~\eqref{eq:tauanalyt}. Improved accuracy of estimating the drag by employing the above equation will be discussed in section~\ref{sec:couetresults}.
\subsection{Results}
\label{sec:couetresults}
In this section, we present the results of the microscale problem as well as the comparison between DNS and the effective model, with emphasis on the prediction of drag components on rough surfaces. We consider square- and ellipse-shaped roughness elements, as shown in figure~\ref{fig:couetmicroprob}(b). Dimensions of both geometries are given in table~\ref{tab:miccouet}. All the simulations presented in this paper are performed using Freefem++\cite{freefem}, an open-source finite element package.
\subsubsection{Microscale problem:}
\begin{table*}
\caption{\label{tab:miccouet} Slip length and stress correction factors for roughness geometries considered in Couette flow.}
\begin{ruledtabular}
\begin{tabular}{ccccccc}
Configuration  & Geometry & $h/l$ &\multicolumn{4}{c}{Constitutive coefficients}\\ \cline{4-7}
  & {Details} &  & $\mathcal{L}_s/l$ & $\mathcal{P}_c$ & $\mathcal{S}_c$ & $(\mathcal{P}_c+\mathcal{S}_c)$ \\ \hline
 \multirow{4}{*}{Ellipse}   & &0.0 &0.06 &0.4972 &0.5042 &1.0014 \\  
                   & {$r_x=0.3l$}  &0.1       &0.16 &0.4971 &0.5026 &0.9998 \\ 
                   &{$r_y=0.6l$}   &0.2      &0.26 &0.4971 &0.5026 &0.9998 \\
                   &                      &0.3     &0.35 &0.4971 &0.5026  &0.9997 \\ \hline
  \multirow{4}{*}{Square} &                      &0.05   &0.0679 &0.3549 &0.6352 & 0.9901 \\
                                           & {$s=0.5l$}   &0.1     &0.1179  &0.3546& 0.6366 &  0.9912\\ 
                                           &                      &0.2    &0.2179  &0.3543 &0.6350 & 0.9894  \\
                                           &                      &0.3    &0.3179 &0.3570  & 0.6322 & 0.9892 \\
\end{tabular}
\end{ruledtabular}
\end{table*}
This section reports the results of the microscale problem presented in equation~\eqref{eqn:couetLprob}. Solution of the microscale problem yields fields of $L_{xx}$, $L_{yx}$, and $B_x$, which can be averaged to compute all three constitutive parameters: $\mathcal{L}_s$, $\mathcal{P}_c$, and $\mathcal{S}_c$ as given in equations~\eqref{eq:Ls_couet} and \eqref{eq:scfactors_couet}. We calculate these parameters for a range of interface heights, and the computed values are presented in table~\ref{tab:miccouet}.  For square roughness elements, interface height $h/l=0$ is not considered. This is because the interface overlaps with a portion of the roughness element in this case, and hence the force balance argument for stress correction factors, presented earlier,  is not directly applicable.  Moreover, in the overlapping portion,  the microscale problem requires simultaneous enforcement of no-slip and stress-jump conditions.  Hence, this is not a well-posed problem.

Data presented in table~\ref{tab:miccouet} show that the slip length~$\mathcal{L}_s$ increases with increasing interface height because the resistance offered by the surface decreases as we move further away. This is consistent with the existing studies~\cite{lacis2020,sudhakar2021}. In contrast, we observe that the values of $\mathcal{P}_c$ and $\mathcal{S}_c$, the constitutive parameters relevant to force prediction, are independent of the chosen interface height.  This is due to the fact that we are solving the Stokes equations in the microscale problem. In the absence of inertial effects, the relative contribution of pressure and viscous effects to the total drag is unaffected by the local flow conditions. Moreover, ignoring the numerical errors,  we can confirm that $\mathcal{P}_c+\mathcal{S}_c=1$.  This observation also indicates that while the contribution of pressure and viscous drag itself might change when moving from smooth to rough surfaces, under the assumption of low Reynolds numbers, the total drag force on the surface remains the same.

\subsubsection{Comparison between DNS and effective model} We demonstrate the effectiveness of the proposed stress correction factors in predicting pressure and viscous drag. The reference data are obtained by performing geometry-resolved simulations~(DNS) over the configuration shown in figure~\ref{fig:couetgeom}(a). We conduct DNS for different Reynolds numbers, $Re=U_wH/\nu$, where $U_w$ and $H$ are defined in figure~\ref{fig:couetgeom}(a), and $\nu$ is the kinematic viscosity of the fluid. We consider the scale separation parameter, $\eta (=l/H) = 0.1$, and the interface height, $h=0.1l$.
\begin{figure*}
\subfloat[]{\includegraphics[scale=0.32]{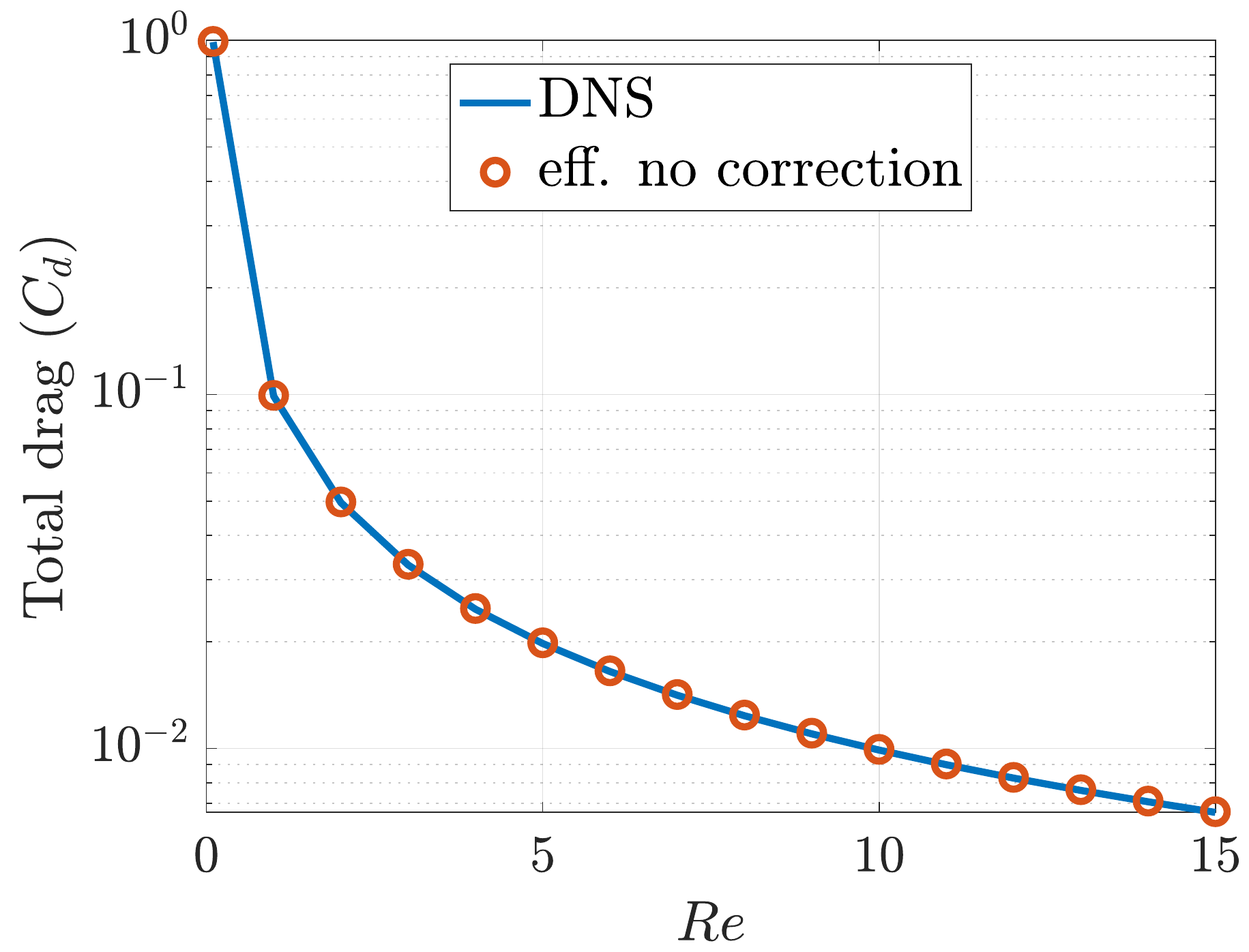}}
\subfloat[]{\includegraphics[scale=0.32]{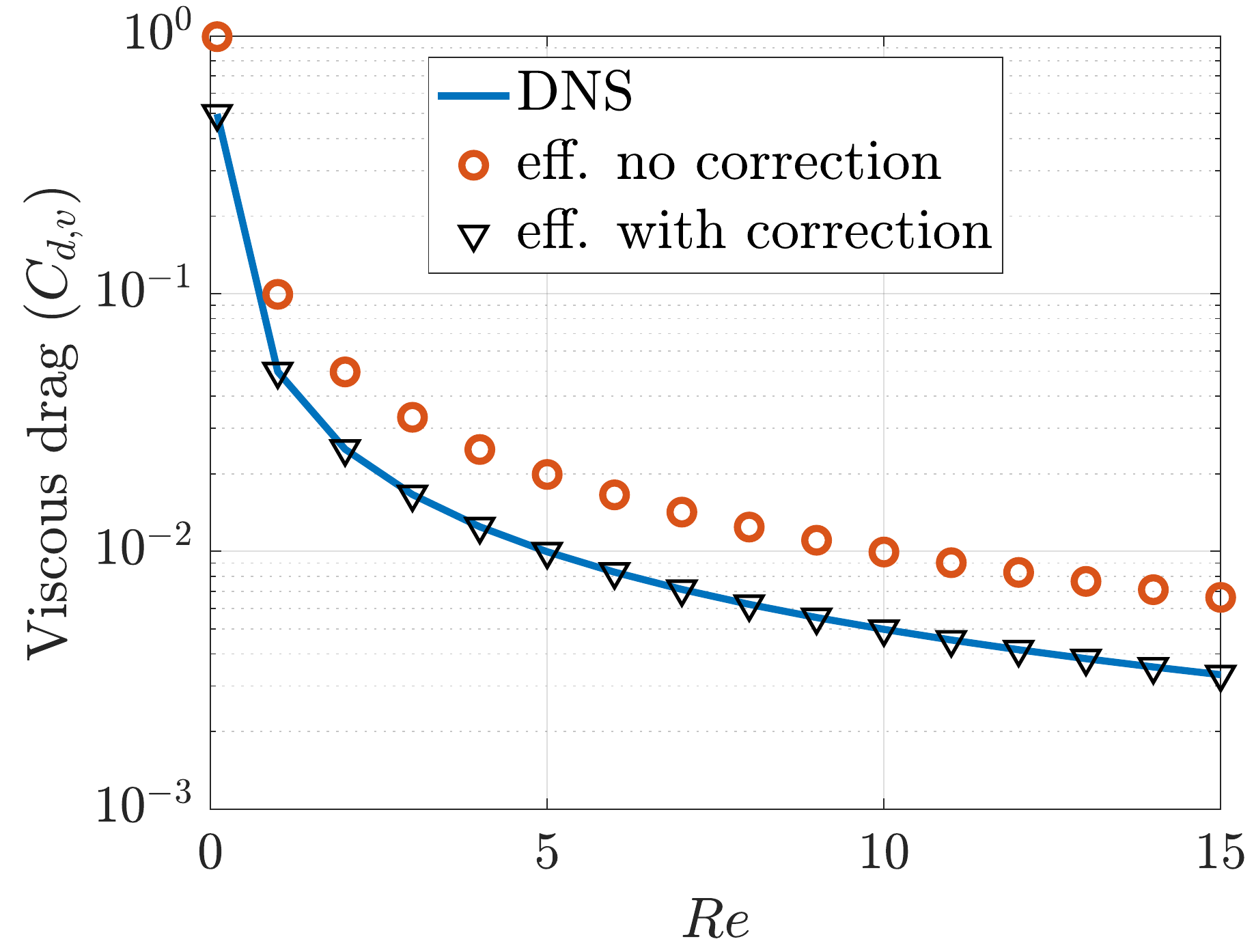}}
\subfloat[]{\includegraphics[scale=0.32]{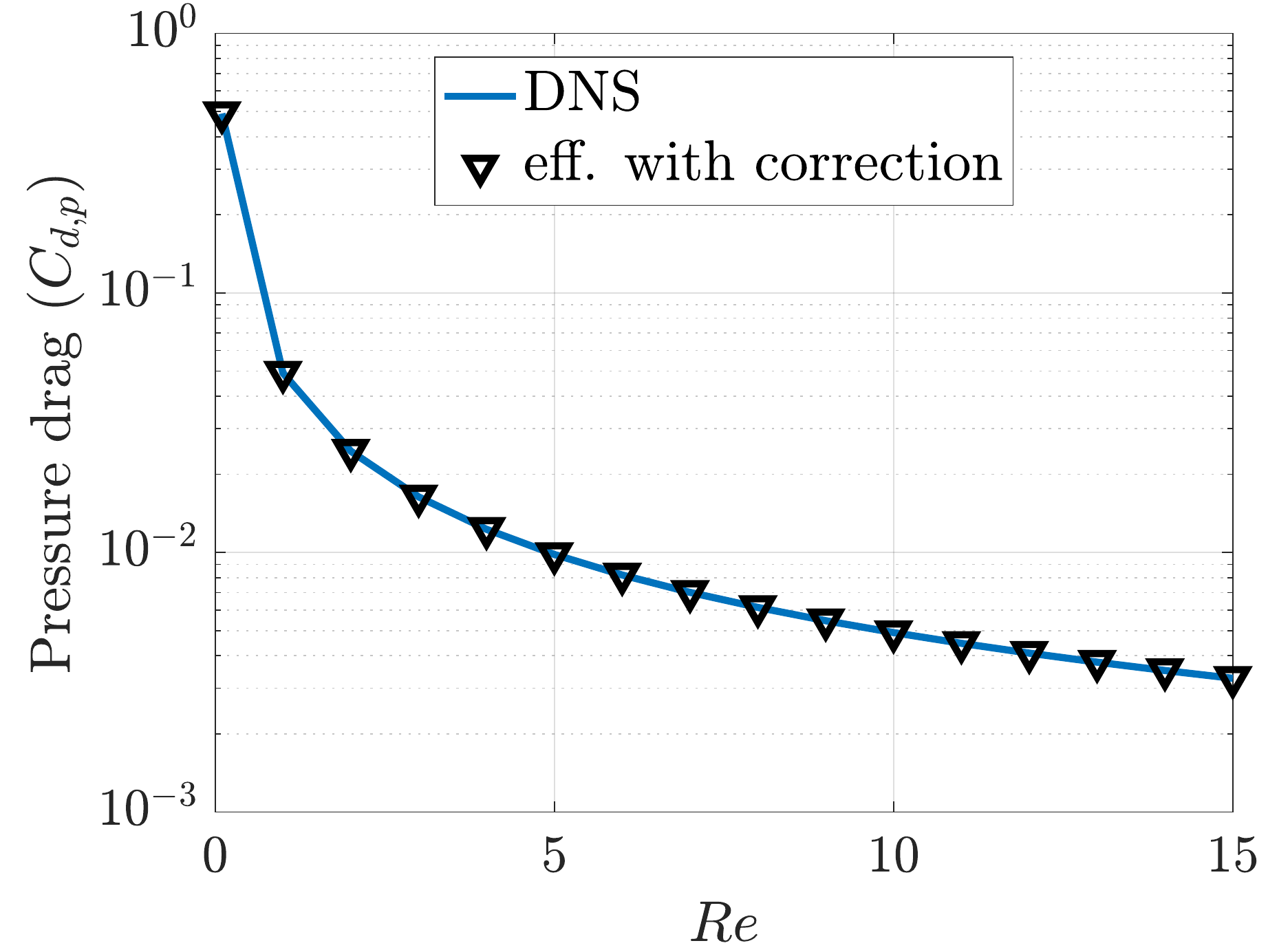}}
\caption{Prediction of drag for Couette flow with elliptic roughness elements for interface height $h=0.1l$. (a)~Total drag, (b)~Viscous drag, and (c)~Pressure drag. The pressure drag predicted using effective models without correction is always zero, leading to 100\% error.}
\label{fig:resdrag}
\end{figure*}

\begin{table*}
\caption{Error in computing slip velocity at the interface, viscous and pressure drag for the Couette flow with a rough wall.}
\label{tab:couetdragres}
\begin{ruledtabular}
\begin{tabular}{ccccccc}
Configuration & $Re$ & \multicolumn{5}{c}{Error ($\%$)} \\ \cline{3-7}
& & & \multicolumn{2}{c}{no correction factors} & \multicolumn{2}{c}{with stress correction factors} \\ \cline{4-5} \cline{6-7}
& & $u_x$ at $\Gamma$ & $C_{d,v}$ & $C_{d,p}$ & $C_{d,v}$ & $C_{d,p}$ \\ \hline
\multirow{6}{*}{Ellipse} & 0.1 & 0.43 & 99.52 & 100 & 0.273  & 0.597 \\
                        			  & 3     & 0.43 & 99.52 & 100 & 0.275  & 0.596 \\
                        			  & 9     & 0.43 & 99.52 & 100 & 0.275  & 0.596 \\
                       			  & 15   & 0.43 & 99.52 & 100 & 0.275  & 0.596 \\ \hline
\multirow{6}{*}{Square}     & 0.1 & 0.08 & 54.21 & 100 & 2.04 & 4.63 \\
				                           & 3   & 0.08 & 54.21 & 100 & 2.04 & 4.63 \\
                           				& 9   & 0.08 & 54.21 & 100 & 2.04 & 4.63  \\
                           			   & 15  & 0.08  & 54.21 & 100 & 2.04 & 4.63 \\
\end{tabular}
\end{ruledtabular}
\end{table*}


\begin{table*}
\caption{Error in computing slip velocity at the interface, viscous and pressure drag for the plane Poiseuille flow with a rough wall.}
\label{tab:poisdragres}
\begin{ruledtabular}
\begin{tabular}{ccccccc}
Configuration & $Re$ & \multicolumn{5}{c}{Error ($\%$)} \\ \cline{3-7}
& & & \multicolumn{2}{c}{no correction factors} & \multicolumn{2}{c}{with stress correction factors} \\ \cline{4-5} \cline{6-7}
& & $u_x$ at $\Gamma$ & $C_{d,v}$ & $C_{d,p}$ & $C_{d,v}$ & $C_{d,p}$ \\ \hline
\multirow{6}{*}{Ellipse} & 0.1 & 2.25& 90.96  & 100 & 4.03  & 10.73 \\
                        			  & 3     & 2.25 & 90.97 & 100  & 4.02  & 10.73 \\
                        			  & 9     & 2.25 & 90.97 & 100  & 4.02  & 10.73 \\
                       			  & 15   & 2.25 & 90.97 & 100  & 4.02  & 10.73 \\ \hline
\multirow{6}{*}{Square}     & 0.1 & 1.30 & 50.80 & 100 & 4.21 & 10.28 \\
				                           & 3   &  1.30 & 50.80 & 100 & 4.21 & 10.28 \\
                           				& 9   &  1.30 & 50.80 & 100 & 4.21 & 10.28  \\
                           			   & 15  &  1.30 & 50.80 & 100 & 4.21 & 10.28 \\
\end{tabular}
\end{ruledtabular}
\end{table*}

Slip velocity obtained from DNS is averaged along the $x-$direction to enable comparison with the effective model. As can be seen from table~\ref{tab:couetdragres}, the model captures tangential velocity at the interface very accurately. This aspect has been addressed extensively in the literature. 

Despite exhibiting high accuracy for the interface velocity, the effective model, without employing stress correction factors, shows a poor prediction of drag components.  The viscous drag coefficient is defined as,
\begin{equation}
    C_{d,v}=\frac{F_v}{\rho U_w^2H}
\end{equation}
where $F_v$ is the viscous drag acting on a single rough element and $\rho$ is the density of the fluid. We computed total viscous drag on the entire rough surface and divided it by the number of roughness elements.  The pressure drag coefficient~($C_{d,p}$) is defined similarly. 

The total, viscous, and pressure drag coefficients on the rough surface for the considered $Re$ are plotted in figure~\ref{fig:resdrag}. 
While the effective model correctly predicts the total drag without the stress correction factors, significant errors are produced in computing the components. The model always predicts zero pressure drag because the roughness elements are cut-off from the simulation domain, and boundary conditions are applied at a fictitious flat surface which is aligned along the flow direction. 
 Such a discrepancy in drag prediction is also reported on flow over a sphere~\cite{zampogna2019}. 
 
Using the modified drag prediction formulae given in equation~\eqref{eq:dragcouet}, we are able to get accurate prediction of both viscous and pressure drag. This is directly evident from figure~\ref{fig:resdrag} and table~\ref{tab:couetdragres}. The reason for the improved performance is the following. The stress correction factors, introduced in the present work, extract information about how the total drag is partitioned into viscous and pressure components from the microscale problem. By employing equation~\eqref{eq:dragcouet}, we use this information and accurately get the components of drag on the rough surface. 

For the Couette flow over rough surfaces, as can be seen from table~\ref{tab:couetdragres} that the error in the prediction of drag components is less than 5\% after applying the stress correction factors.  However, the Couette flow represents an ideal test case because the velocity profile across the channel height is linear, and the shear stress is independent of the interface location~($y_i$). 

In order to test the effectiveness of the proposed correction factors for other scenarios, we simulate plane Poiseuille flow within the configuration given in figure~\ref{fig:couetgeom}: we set $U_w=0$ and apply a body force along the $x-$direction to drive the flow. The Reynolds number is defined as $Re=U_{avg}H/\nu$, where $U_{avg}$ is the average velocity. Other details of DNS and the microscale problem remain the same as that of the Couette flow.

Errors introduced by the effective model in the prediction of interface slip velocity and drag components are presented in table~\ref{tab:poisdragres}. Although the error introduced is larger than that of Couette flow, it is directly evident that the proposed stress correction factors enable us to compute both viscous and pressure drag more accurately than those without these factors. This observation is consistent in both Couette and plane Poiseuille flows.

In order to prove the consistency of our stress correction factors, we vary the interface height~($h$) in the Poiseuille flow simulations and compute the error produced in the computation of drag components. We have performed this test on elliptical roughness at $Re=100$, and the results are presented in Table~\ref{tab:poisdragyi}. We can see that the error consistently increases with moving the interface location away from the rough surface. This behaviour is similar to those reported in the literature~\cite{lacis2020,sudhakar2021}, and this shows consistency of stress correction factors.
\begin{table*}
\caption{Influence of interface height in computing drag coefficients for the Poiseuille flow with elliptic roughness elements at $Re=100$.  Drag coefficients,  non-dimensionalised using section-averaged velocity, obtained from DNS are $C_{d,v}=0.003$ and $C_{d,p}=0.0032$.}
\label{tab:poisdragyi}
\begin{ruledtabular}
\begin{tabular}{ccc}
$h/l$ &  \multicolumn{2}{c}{Error ($\%$)} \\ \cline{2-3}
& $C_{d,v}$ & $C_{d,p}$  \\ \hline
0.0	 &  1.64  &  9.02 \\
0.05  &  2.94  &  9.92 \\
0.1    &  4.02  &  10.73 \\
0.2   &  5.90  &  12.67  \\
0.3   &  7.88  &  14.50 \\
0.4   &  9,83  &  16.33 \\
0.5   &  11.79  &  18.16 \\
\end{tabular}
\end{ruledtabular}
\end{table*}
\section{Flow over rough cylinders}
\label{sec:cyl}
The previous section shows that the stress correction factors introduced in this paper lead to the accurate prediction of pressure and viscous drag. However, the fictitious interface in the previous section was aligned along one of the coordinate axes. To investigate the effectiveness of the proposed stress correction factors in a complex problem, we consider here the flow over a rough circular cylinder. In order to solve the problem, we formulate the TR model in polar coordinates. We examine the effectiveness of the model in predicting the interface velocity before addressing the accuracy of drag prediction.
\begin{figure*}
\subfloat[]{\includegraphics[scale=0.7]{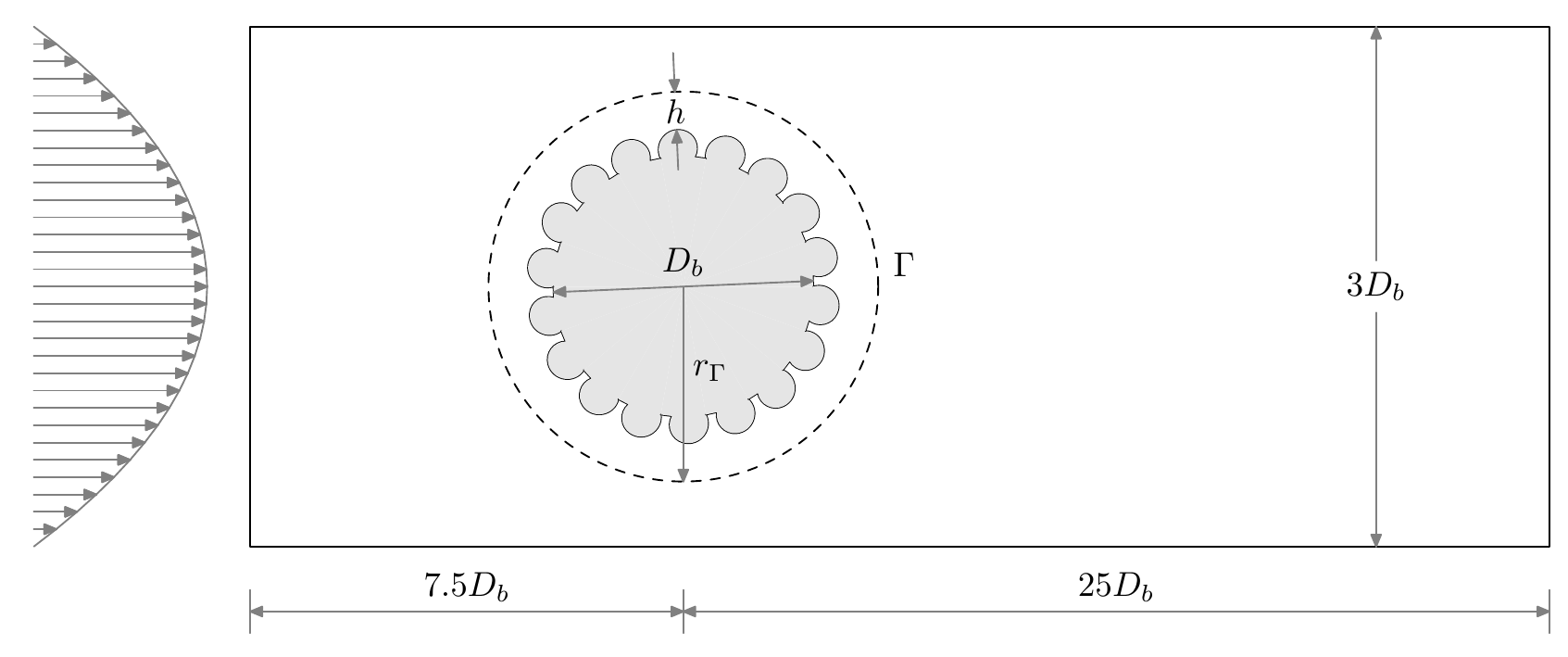}}\\
\subfloat[]{\includegraphics[scale=0.85]{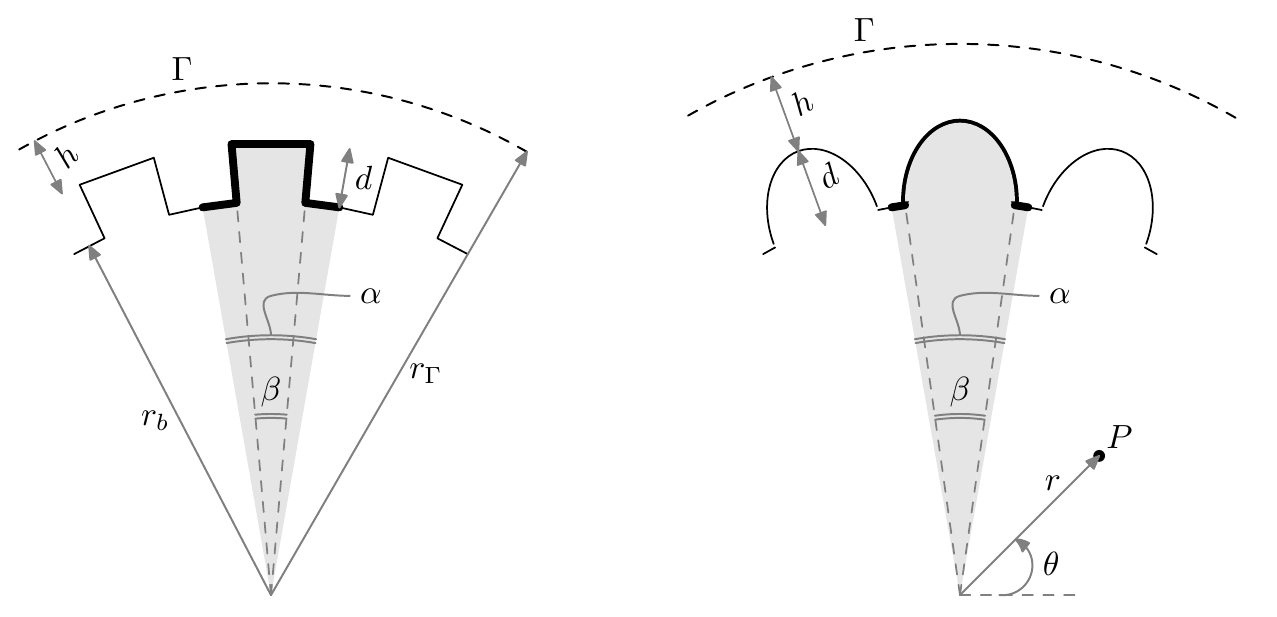}}
\caption{Flow over rough cylinders. (a)~Computational domain (not to scale).  The origin of the polar coordinates is at the centre of the circle, (b)~Geometric details of roughness elements.}
\label{fig:cyldomain}
\end{figure*}

The configuration of the problem is shown in figure~\ref{fig:cyldomain}(a).  $D_b$ denotes the base diameter on which ordered roughness elements are added.  The roughness height, measured along the radial direction, is $d$.  We consider two types of roughness geometries, square- and ellipse-shaped, as shown in figure~\ref{fig:cyldomain}(b).  

No-slip condition is enforced on the top and bottom boundaries; a parabolic profile for velocity is specified on the left boundary.  Zero-stress condition is applied on the right boundary, as shown in figure~\ref{fig:cyldomain}(a). 

The Reynolds number is defined as $Re=U_\textrm{max} D_b/\nu$, where $U_\textrm{max} $ is the maximum velocity at the inlet,  and $\nu$ is the kinematic viscosity of the fluid. The parameters used in the simulations are as follows: $D_b=20$, $d=1$, $\alpha=10\deg$, and $\beta$ is set as $5\deg$ and $8\deg$ for square and elliptic roughness geometries, respectively. $\nu$ is adjusted to set $Re$.  $d$ and $D_b$ are chosen to be the relevant length scales for microscopic and macroscopic effects. This leads to the scale separation parameter, $\eta=d/D_b=0.05$.

As shown in tables~\ref{tab:miccyl} and \ref{tab:cyldragres}, we perform simulations for three different $Re$ and compare results of the effective model against DNS for different interface heights~($h$). 
\subsection{Interface velocity formulation}
The effective model replaces the rough cylinder with a smooth fictitious circular interface placed at a distance of $h$ above the crest plane of roughness. This section provides effective interface conditions in polar coordinates. As shown in figure~\ref{fig:cyldomain}(b), $r$ and $\theta$ represent radial and transverse direction, respectively; $u_r$ and $u_\theta$ denote velocity components along these directions. 

The interface condition for the transverse velocity component can be directly written from the vectorial boundary condition formulations presented  by Ugis et  al.\cite{lacis2020}. It reads 
\begin{equation}
u_\theta = \mathcal{\bm{L}}\bm{\varepsilon}\cdot\bm{n}
\end{equation}
where $\mathcal{\bm{L}}=\{\mathcal{L}_{\theta r},\mathcal{L}_{\theta\theta}\}$ is the slip length vector,  $\bm{n}$ is the unit normal vector to the interface, and $\bm{\varepsilon}$ is the rate of strain tensor
\begin{equation}
\bm{\varepsilon}=
\begin{bmatrix}
\frac{\partial u_r}{\partial r} & \frac{1}{2}\left(r\frac{\partial}{\partial r}\left( \frac{u_\theta}{r}\right)+\frac{1}{r}\frac{\partial u_r}{\partial \theta}\right)\\
\frac{1}{2}\left(r\frac{\partial}{\partial r}\left( \frac{u_\theta}{r}\right)+\frac{1}{r}\frac{\partial u_r}{\partial \theta}\right) & \frac{1}{r}\frac{\partial u_\theta}{\partial \theta}+\frac{u_r}{r}
\end{bmatrix}
\end{equation}

We can infer from arguments presented in the earlier works\cite{lacis2016,lacis2020,sudhakar2021} that  $\mathcal{L}_{\theta r}=0$. Hence the condition for the transverse velocity can be written as,
\begin{equation}
u_\theta=\mathcal{L}_{\theta\theta}\frac{1}{2}\left[r\frac{\partial}{\partial r}\left(\frac{u_\theta}{r}\right)+\frac{1}{r}\frac{\partial u_r}{\partial \theta}\right]
\label{eqn:polut1}
\end{equation}

\begin{figure*}
\includegraphics[scale=0.2]{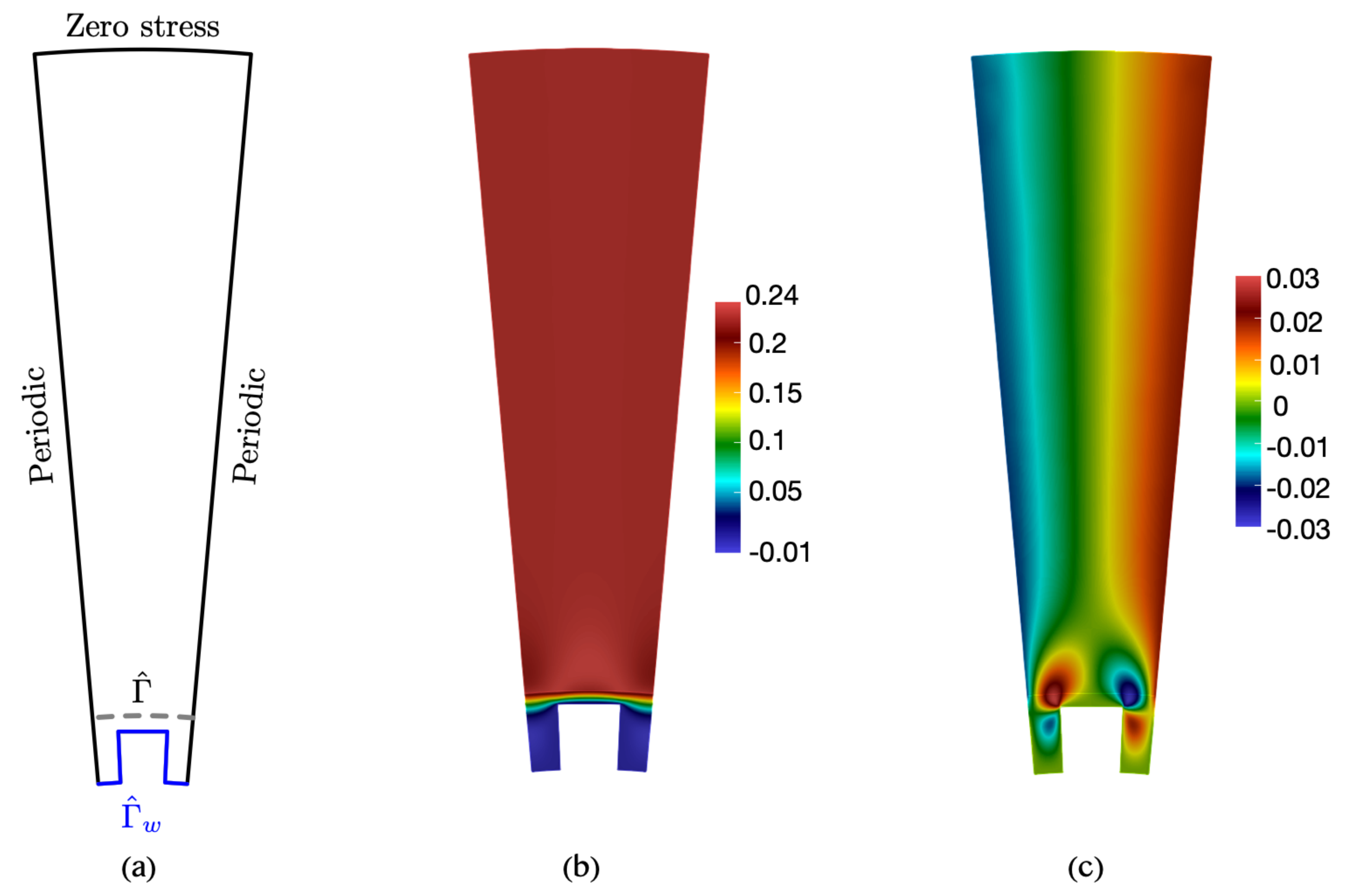}
\caption{Microscale problem in polar coordinates. for square roughess elements with the interface height $h=0.2$ (a)~Boundary conditions, (b)~field of $L_{\theta\theta}$,  and (c)~field of $L_{r\theta}$.}
\label{fig:cylmicroprob}
\end{figure*}

Similar to the formulations presented for the Cartesian coordinate system, $\mathcal{L}_{\theta\theta}$ can be computed by solving a microscale problem on an interface cell. The geometry of the interface cell is shown in figure~\ref{fig:cylmicroprob}(a). Additional details are presented in section~\ref{sec:constpolar}.

The governing equations for the microscale problem based on discussions in Ugis et al.\cite{lacis2020}  can be written as,
 \begin{subequations}
 \begin{align}
     \nabla\cdot L^{\pm}_{i\theta}&=0 \quad \textrm{on } \hat{\Omega}^{\pm}\\
     -\nabla B^{\pm}_\theta + \Delta L^{\pm}_{i\theta} &=0 \quad \textrm{on } \hat{\Omega}^{\pm}\\
     \llbracket L_{i\theta} \rrbracket &= 0 \quad \textrm{on } \hat{\Gamma}\\
     \left\llbracket -B_\theta+2\frac{\partial L_{r\theta}}{\partial r} \right\rrbracket &= 0 \quad \textrm{on } \hat{\Gamma}\\
     \left\llbracket r\frac{\partial}{\partial r}\left( \frac{u_\theta}{r}\right)+\frac{1}{r}\frac{\partial u_r}{\partial \theta}\right\rrbracket &= -1 \quad \textrm{on } \hat{\Gamma} \label{eqn:polLprob_e}
 \end{align}
 \label{eqn:polLprob}
 \end{subequations}
 where $i\in \{r,\theta\}$.  As mentioned earlier,  quantities with superscripts $+$ and $-$ indicate fields above and below the interface respectively. 
 
 Solution of the above microscale problem yields fields of $L_{r\theta}$ and $L_{\theta\theta}$ in the microscale domain. The relevant constitutive parameter is calculated by the following averaging carried out at the interface
 \begin{equation}
 \mathcal{L}_{\theta\theta}=\frac{1}{\alpha}\int_{\theta_1}^{\theta_2}{L_{\theta\theta}d\theta}.
 \label{eqn:polavgi}
 \end{equation}
 
The microscale problem~(equation~\eqref{eqn:polLprob}) and the averaging formula~(equation~\eqref{eqn:polavgi}) complements the interface condition for $u_\theta$ (equation~\eqref{eqn:polut1}). However, for a complete specification of the problem,  a formulation for $u_r$ should also be supplied.  Most studies on effective models for rough walls assume wall normal velocity,  $u_r$ in polar coordinates, is zero at the interface. However, as shown by Ugis et al.\cite{lacis2020}, this is true only when the tangential velocity component ($u_\theta$ in our case) is constant along the entire interface length.  When $u_\theta$ varies along $\Gamma$, mass conservation induces a non-zero $u_r$.

\begin{figure}
\includegraphics[scale=0.9]{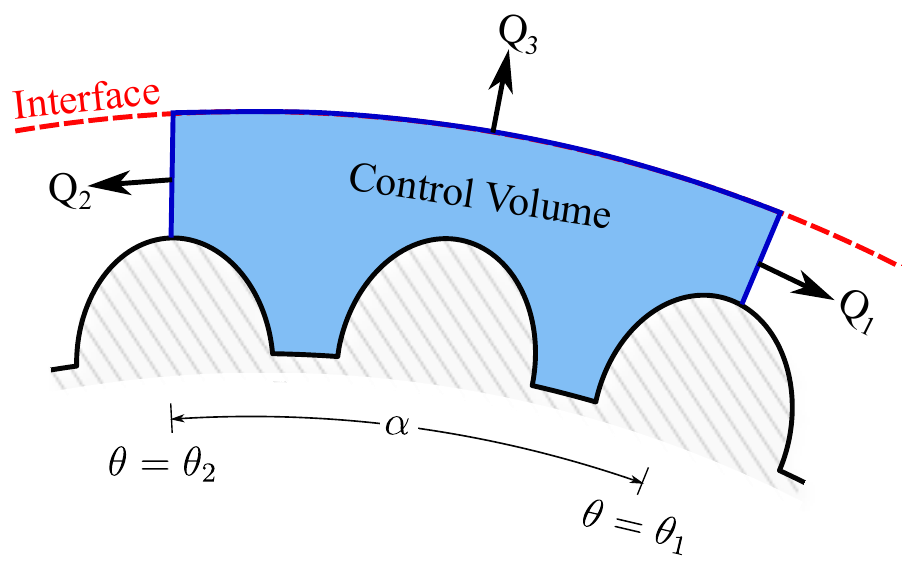}
\caption{Control volume in polar coordinates around rough elements.}
\label{fig:polcv}
\end{figure}

Following the procedure explained by Ugis et  al.\cite{lacis2020},  by applying the mass conservation over a control volume,  we derive the formulation for radial velocity component at the interface.   Consider the control volume as shown in figure~\ref{fig:polcv}.  Mass conservation requires
\begin{equation}
\sum_i{Q_i} = 0,
\label{eqn:massint}
\end{equation}
where $Q_i$ is volume flux through control surface $i$.  Flux through face 1 can be written as
\begin{equation}
Q_1=\int_{r_b}^{r_\Gamma}{\langle u \rangle(\hat{r})\cdot\bm{n}d\hat{r}},
\end{equation}
where $\langle u \rangle(\hat{r})$ is velocity that is averaged along the transverse direction at a given $\hat{r}$, as defined below
\begin{equation}
\langle u \rangle(\hat{r}=a)=\frac{1}{\alpha}\int_{\theta_1}^{\theta_2}{u(\hat{r}=a,\hat{\theta})\ d\hat{\theta}}
\end{equation}
Substituting the Ansatz for $u_\theta$\cite{lacis2020},  equation for $Q_1$ can be written as
\begin{equation}
Q_1=-\int_{r_b}^{r_\Gamma}{\langle L^-_{\theta\theta} \rangle(\hat{r})\varepsilon_{r\theta}d\hat{r}}.
\end{equation}
Using the definition of $\langle L^-_{\theta\theta} \rangle$ and noting that $\varepsilon_{r\theta}$ is independent of microscale coordinates, we can write
\begin{equation}
Q_1=-\varepsilon_{r\theta}\vert_{\theta=\theta_1}\mathcal{N},
\end{equation}
where
\begin{equation}
\mathcal{N}=\frac{1}{\alpha}\int_{r_b}^{r_\Gamma}{\int_{\theta_1}^{\theta_2}{\left<L^-_{\theta\theta}\right>(\hat{r})d\hat{r}}d\hat{\theta}} = \frac{1}{\alpha}\int_{\hat{\Omega}^-}{\left(\frac{L^-_{\theta\theta}}{\hat{r}}.\right) d\hat{\Omega}^-}
\end{equation}
Similarly we can write
\begin{equation}
Q_2=\varepsilon_{r\theta}\vert_{\theta=\theta_2}\mathcal{N}.
\end{equation}
Volume flux through face 3 can be written as
\begin{equation}
Q_3=r_\Gamma\int_{\theta_1}^{\theta_2}{u_r d\hat{\theta}}.
\end{equation}
Substituting $Q_1$, $Q_2$ and $Q_3$ in equation~\eqref{eqn:massint}, we get
\begin{equation}
r_\Gamma\int_{\theta_1}^{\theta_2}{u_r d\hat{\theta}}=\left[\varepsilon_{r\theta}\vert_{\theta=\theta_1}-\varepsilon_{r\theta}\vert_{\theta=\theta_2}\right]\mathcal{N}.
\end{equation}
Shrinking the control volume to be infinitesimally small $\alpha=\Delta \theta\to 0$, and making use the definition of the derivative,
\begin{equation}
u_r = -\frac{\mathcal{N}}{r_{\Gamma}}\frac{d\epsilon_{r\theta}}{d\theta}.
\end{equation}
Using equation~\eqref{eqn:polut1},  we can rewrite the above expression as
\begin{equation}
u_r = -\mathcal{M}\frac{du_{\theta}}{d\theta},
\end{equation}
where
\begin{equation}
\mathcal{M}=\frac{1}{\alpha r_\Gamma\mathcal{L}_{\theta\theta}}\int_{\hat{\Omega}^-}{\left(\frac{L^-_{\theta\theta}}{\hat{r}}.\right) d\hat{\Omega}^-}.
\label{eqn:polarm}
\end{equation} 

In summary, the interface velocity conditions, in the framework of TR model, in polar coordinates can be written as
\begin{subequations}
\begin{align}
u_\theta&=\mathcal{L}_{\theta\theta}\frac{1}{2}\left[r\frac{\partial}{\partial r}\left(\frac{u_\theta}{r}\right)+\frac{1}{r}\frac{\partial u_r}{\partial \theta}\right]\\
u_r&=-\mathcal{M}\frac{\partial u_\theta}{\partial \theta}
\end{align}
\label{eqn:polvelo}
\end{subequations}
For the problem under consideration,  $u_\theta$  and $u_r$  represent the tangential and the transpiration velocity components at the interface respectively.

\subsection{Computation of constitutive coefficients}
\label{sec:constpolar}
The microscale problem presented in equation~\eqref{eqn:polLprob} is solved on the interface cell (figure~\ref{fig:cylmicroprob}(a)) to compute all four constitutive coefficients: $\mathcal{L}_{\theta\theta}$, $\mathcal{M}$, $\mathcal{P}_c$ and $\mathcal{S}_c$. Periodic boundary condition is applied on radial boundaries, no-slip is enforced on the cylinder surface, and zero stress condition is specified on the outer boundary, as shown in figure~\ref{fig:cylmicroprob}(a). Similar to the microscale problem in the Cartesian coordinates, the flow is driven by interface forcing, as shown in equation~\eqref{eqn:polLprob_e}.

\begin{table*}
\caption{\label{tab:miccyl} Constitutive parameters for the rough cylinder problem.} 
\begin{ruledtabular}
\begin{tabular}{cccccc}
Configuration  & $h$ &\multicolumn{4}{c}{Constitutive coefficients}\\ \cline{3-6}
&  & $\mathcal{L}_{\theta\theta}$ & $\mathcal{M}$ & $\mathcal{P}_c$ & $\mathcal{S}_c$  \\ \hline
 \multirow{4}{*}{Ellipse} & 0 &0.1030 & 0.00982 &0.3452 & 0.6515  \\
									 & 0.05 &0.1446 &0.00984 &0.3805 & 0.6178  \\
                                         &0.1    &0.1943 &0.01118  &0.3816  & 0.6166  \\
                   					&0.2    &0.2944 &0.01471 &0.3834 &0.6147 \\
                   				    &0.3   &0.3956 &0.01861 &0.3848  &0.6133  \\ \hline
 \multirow{4}{*}{Square} & 0.05 &0.0834 & 0.00479 &0.3414 &0.6445 \\
                                         &0.1    &0.1331   &0.00659  &0.3408 &0.6453  \\
                   					&0.2    &0.2332  &0.01070  &0.3391 &0.6476 \\
                   				    &0.3   &0.3343  &0.01494  &0.3375 &0.6488  \\
\end{tabular}
\end{ruledtabular}
\end{table*}

The computational domain for the microscale problem, shown in figure~\ref{fig:cylmicroprob}(a), contains one roughness element along the transverse direction. In the radial direction, it extends from physical roughness elements to the outer boundary, which is located at a radial distance of $5\left|\hat{\Gamma}\right|$ from the interface. Increasing the distance of the outer radial boundary from the interface to $10\left|\hat{\Gamma}\right|$ does not cause any noticeable difference in the value of constitutive parameters.

Solution of the microscale problem yields fileds of $L_{r\theta}$ and $L_{\theta\theta}$ in the interface cell.  These are presented for square roughness elements with interface height $h=0.2$ in figure~\ref{fig:cylmicroprob}. Constitutive parameters $\mathcal{L}_{\theta\theta}$ and $\mathcal{M}$ are computed from these fields using equations~\eqref{eqn:polavgi} and \eqref{eqn:polarm}. For the two microscale roughness geometries considered in this work, the constitutive coefficients for different interface heights are listed in Table~\ref{tab:miccyl}. As expected, with increasing interface height, both $\mathcal{L}_{\theta\theta}$ and $\mathcal{M}$ increase in magnitude.

By replacing $L_{ix}$ and $B_x$ with $L_{i\theta}$ and $B_\theta$ respectively,  we can compute stress correction factors using equation~\eqref{eq:scfactors_couet}. It can be inferred from values presented in Table~\ref{tab:miccyl} that $\mathcal{P}_c$ and $\mathcal{S}_c$ values are independent of interface height. Also, based on the force balance argument presented earlier, neglecting a small error caused by the numerical method, $\mathcal{P}_c+\mathcal{S}_c=1$.
\subsection{Computation of drag}
The uncorrected drag force can be computed by the usual procedure of integrating the stress along the effective interface using
 \begin{subequations}
 \begin{align}
     F^{uncorr}_v &= \int_{\Gamma}{({\tau}\cdot \bm{n})\cdot \bm{e}_x\ d\Gamma}, \\
     F^{uncorr}_p & = -\int_{\Gamma}{pn_x\ d\Gamma},
 \end{align}
 \label{eq:pundrag}
 \end{subequations}
where $\bm{n}$ is the normal vector to the effective cylinder surface $\Gamma$, and the subscript $x$ denotes its component in the Cartesian coordinate system.  The superscript `uncorr' implies these forces are uncorrected. $\bm{\tau}$ and $\bm{e}_x$ are the deviatoric part of the stress tensor and the unit vector along $x-$direction, respectively.  We will see in the subsequent sections that the direct computation of forces using the above formulations will lead to large errors. These errors, as will be shown, can be significantly reduced by employing the stress correction factors proposed in this work. Components of the drag force, after employing the correction factors,  can be expressed as
 \begin{subequations}
 \begin{align}
     F_v &= \mathcal{S}_c F^{uncorr}_v\\
     F_p & = F^{uncorr}_p+\mathcal{P}_c F^{uncorr}_v
 \end{align}
 \label{eq:poldragcorr}
 \end{subequations}
As seen in the above equation, the pressure drag has two contributions. The first term is associated with the pressure distribution on the effective smooth cylinder surface.  The second term is due to the contribution from microscale geometrical features of the roughness elements.

The non-dimensional pressure and viscous drag coefficients, computed using corrected forces,  can be expressed as
\begin{equation}
C_{d_{\{v,p\}}} = \frac{2F_{\{v,p\}}}{\rho U^2 D_b}
\end{equation}
Similarly, $C_d^{uncorr}$ implies coefficients computed using uncorrected forces given in equation~\eqref{eq:pundrag}.


\subsection{Results}
In order to test the accuracy of the proposed formulations, we compare the results obtained from the effective model against DNS results.  In this section, we report the accuracy of the effective model in predicting interface velocities as well as drag components.

 \begin{figure}
\subfloat[]{\includegraphics[width=8cm]{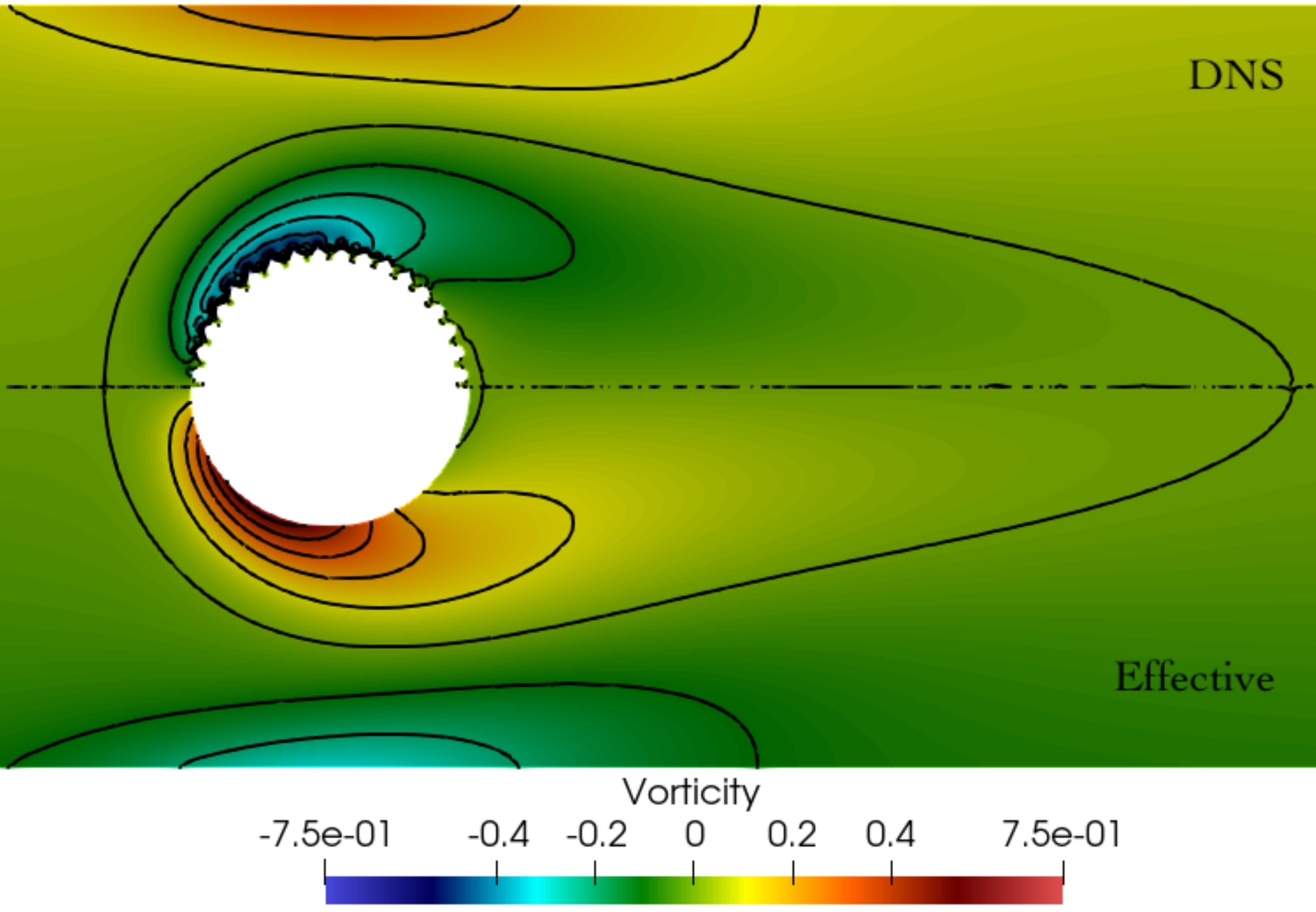}}
\subfloat[]{\includegraphics[trim = 0cm -0.5cm 0cm -0.5cm, width=8cm]{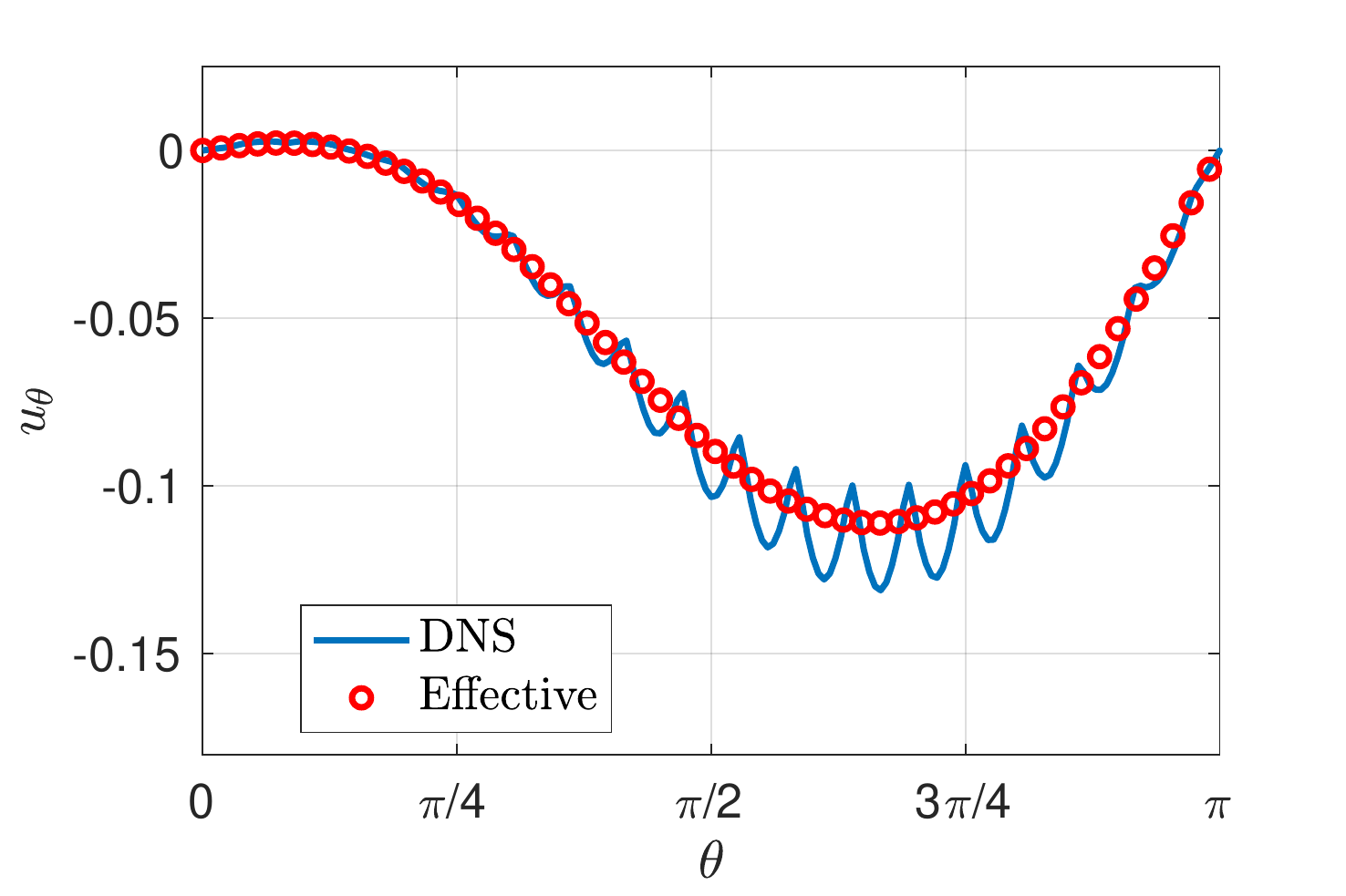}}\hspace{0.5cm}
\caption{Comparison for DNS and effective model for rough cylinder with elliptic roughness elements at $Re=20$. Vorticity contours of DNS (top half) and effective model (bottom half),  (b)~Transverse velocity. Effective model results are taken from the simulation with $h=0.1$. $\theta=0$ corresponds to the rear end of the cylinder.}
\label{fig:instcompare}
\end{figure}

Vorticity contours for flow over the cylinder with elliptic inclusions at $Re=20$ is given in figure~\ref{fig:instcompare}(a).  The top-half of the figure shows results from DNS and the bottom-half contains the vorticity obtained from the effective model.  The plot clearly shows that the effect model captures the flow field accurately, except for the minor variation very close to the rough surface. This difference is due to the fact that homogenisation-based effective models involve an inherent averaging process that eliminates microscale variation of flowfield quantities.  This can be clearly seen from figure~\ref{fig:instcompare}(b) which shows variation of transverse velocity along the cylinder surface.  Due to symmetry,  only the top half of the interface ($0\le\theta\le  \pi$) is considered.  While the effective model accurately captures the macroscale variation, as expected, the microscale oscillations are smoothed out. Hence,  by construction,  effective models, can predict only macroscale quantities and their variation in macroscale.

First, we validate the interface velocity formulations proposed in polar coordinates~(equation~\eqref{eqn:polvelo}).  In order to enable the comparison between DNS and the effective model, DNS results should be smoothed out by eliminating microscale variations. We achieve this by performing $N$ number of DNS and carry out ensembled averaging of these simulations. In each simulation, the rough cylinder is rotated by an angle $\Delta\alpha$ such that $N\cdot \Delta\alpha=\alpha$. This procedure is similar to that used by Sudhakar et al.\cite{sudhakar2021}.

\begin{figure}
\subfloat[]{\includegraphics[scale=0.55]{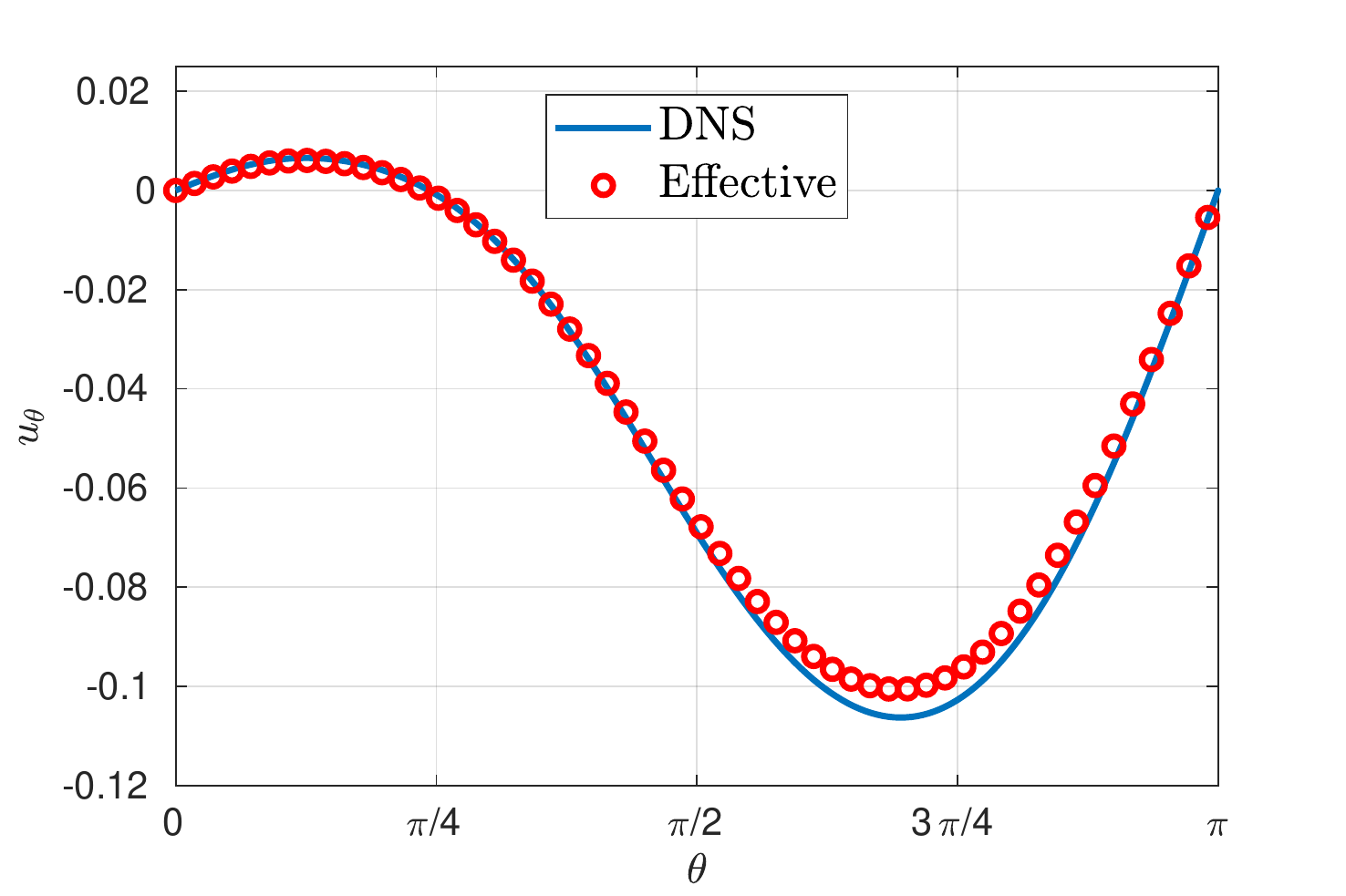}}
\subfloat[]{\includegraphics[scale=0.55]{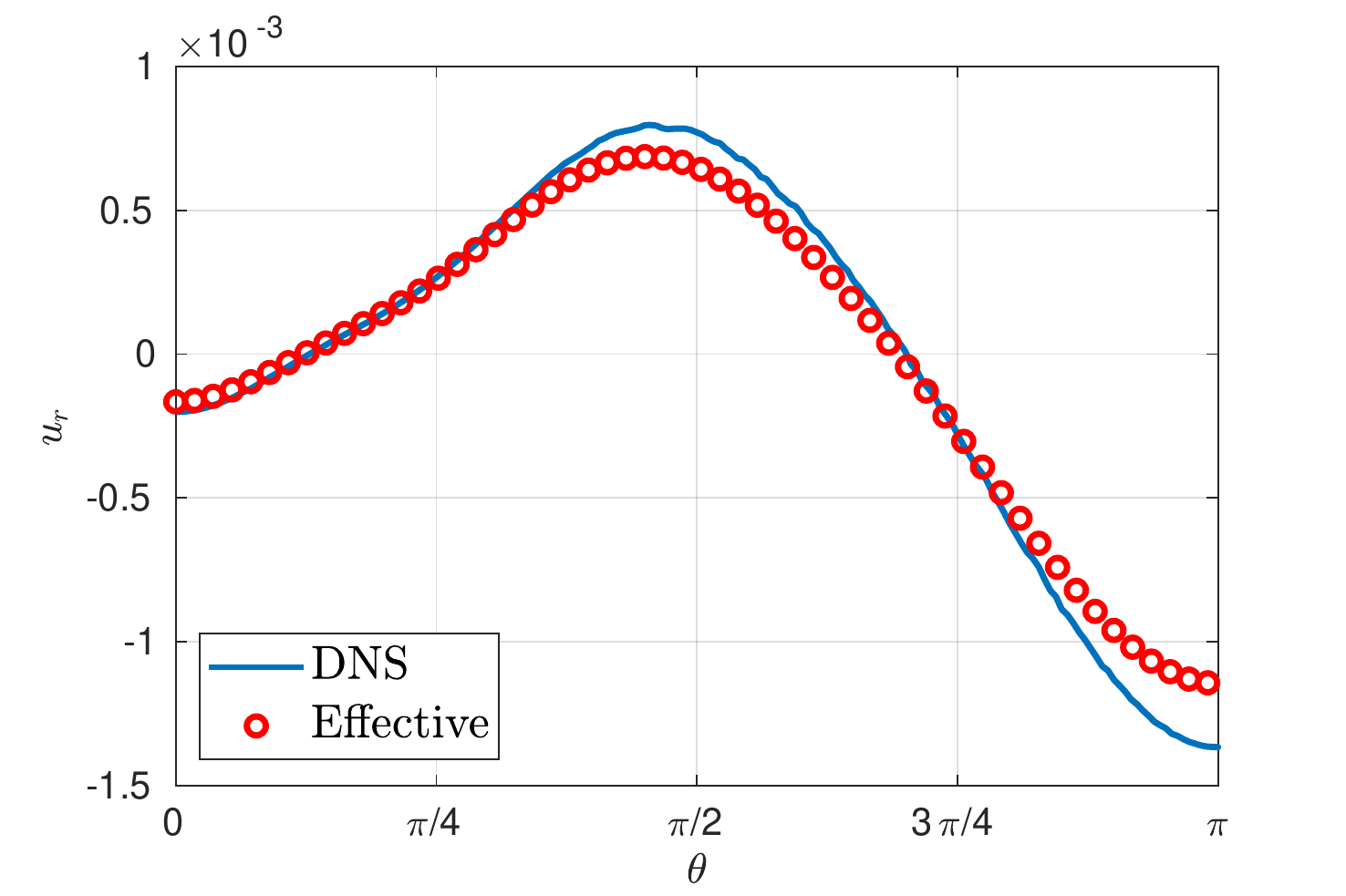}}
\caption{Velocity around the rough cylinder with square roughness elements at $h=0.1$ and $Re=40$.  (a)~Transverse velocity, (b)~Radial velocity. $\theta=0$ corresponds to the rear end of the cylinder.}
\label{fig:polarplot}
\end{figure}

Variation of transverse and radial velocity components along the interface is presented in figure~\ref{fig:polarplot}.  It can be seen that over the entire length of the interface,  results from the effective model match well with DNS results.  In order to quantify the accuracy in the prediction of interface velocity components, we compare errors in minimum transverse velocity~($u_\theta^{min}$) and maximum radial velocity~($u_r^{max}$). Error in quantity $q$ is defined as
\begin{equation}
e_q=\left|\frac{q_\textrm{eff}-q_\textrm{DNS}}{q_\textrm{DNS}}\right|,
\label{eqn:qerr}
\end{equation}
where $q_\textrm{eff}$ and $q_\textrm{DNS}$ denote quantities obtained from the effective model and DNS, respectively.

Errors in $u_\theta^{min}$ and $u_r^{max}$, computed using equation~\eqref{eqn:qerr} are presented in Table~\ref{tab:cyldragres}. For all three $Re$ considered, the error in $u_\theta^{min}$ is less thab 8\%,  and the error in $u_r^{max}$ is within 14\%.  The error increases with increasing $Re$. This is due to the fact that the interface velocity formulations are derived based on the assumption of Stokes flow near the interface, and with increasing $Re$,  the inertial effects become non-negligible.  The above observations are true for both geometry of roughness elements considered.  As can be seen from the table, the model induces a larger error in predicting the transpiration velocity than that of the tangential component.  A similar observation of effective models is reported in the literature\cite{lacis2020,sudhakar2021}. From these observations, we can state that the proposed model provides sufficiently accurate results in predicting interface velocity components. 

\begin{table*}
\caption{Error in computing interface velocity,  and drag components for flow over rough cylinders at $h=0.1$.}
\label{tab:cyldragres}
\begin{ruledtabular}
\begin{tabular}{cccccccc}
Configuration & $Re$ & \multicolumn{6}{c}{Error ($\%$)} \\ \cline{3-8}
& & & & \multicolumn{2}{c}{no correction factors} & \multicolumn{2}{c}{with stress correction factors} \\ \cline{5-6} \cline{7-8}
& & $u_\theta^{min}$ & $u_r^{max}$ & $C_{d,p}$ & $C_{d,v}$ & $C_{d,p}$ & $C_{d,v}$ \\[0.075cm] \hline
\multirow{3}{*}{Ellipse} & 1    & 4.850 & 7.921  & 15.938  & 54.938  & 3.057  & 4.471  \\
 								    & 20 & 6.720 & 12.168 & 14.560  & 52.932  & 1.969  & 5.708  \\
								    & 40 & 7.952 & 13.893 & 12.304 & 52.278 & 1.709   & 6.111  \\ \hline
\multirow{3}{*}{Square} & 1  & 3.216 & 6.370    & 14.722   & 50.672  & 2.514  & 2.676  \\
 								    & 20 & 4.745 & 12.769   & 13.521  & 48.872  & 1.481  & 3.929  \\
								    & 40 & 5.339 & 13.802 & 11.408  & 48.269  & 1.319  & 4.318  \\ 
\end{tabular}
\end{ruledtabular}
\end{table*}

Another important point to note is that although, $u_r$ is an order of magnitude smaller than $u_\theta$,  based on results of Ugis et al.\cite{lacis2020}, we can state that accurately capturing $u_r$ will be essential to produce physically consistent results for turbulent flow over rough cylinders.

After comparing the interface velocities between DNS and the effective model,  we study the accuracy of the model in predicting drag components.  As presented in the Couette flow problem, the effective model without correction factors give accurate results for total drag on rough cylinders.

Percentage of errors in computing drag components,  using equation~\eqref{eqn:qerr}, are reported in table~\ref{tab:cyldragres}. It can be observed that without correction factors,  the errors are very large for both drag components.  In contrast to the Couette/Poissuille flow cases for which the error in pressure drag is always 100\%,  the error in viscous drag is larger in this case.  This has to do with the alignment of the interface with respect to the flow direction.  By incorporating stress correction factors using equation~\eqref{eq:poldragcorr}, we are able to get accurate prediction of both drag components, as evident from table~\ref{tab:cyldragres}. This confirms that the proposed stress correction factors help us determine how the total drag is partitioned into pressure and viscous drag at rough curved surfaces also.

\begin{table*}
\caption{Influence of the interface height in computing interface velocities and drag coefficients for flow over a cylinder with elliptic roughness elements at $Re=40$.  Drag coefficients obtained from DNS are $C_{d,v}=0.5889$ and $C_{d,p}=2.4418$.} 
\label{tab:cyldragyi}
\begin{ruledtabular}
\begin{tabular}{ccccc}
$h$ &  \multicolumn{4}{c}{Error ($\%$)} \\ \cline{2-5}
& $u_\theta^{min}$ & $u_r^{max}$ & $C_{d,v}$ & $C_{d,p}$  \\ \hline
0       &  0.730  & 3.969      &   0.639  &  0.784 \\
0.05  & 6.930  &  14.040    &   5.624  &  1.542 \\
0.1    & 7.592  &  13.893     &  6.111  &  1.709 \\
0.2   & 9.095  &  14.394     &   7.041  &  2.063 \\
0.3  & 10.580  &  15.506     &   7.918  &  2.444 \\
\end{tabular}
\end{ruledtabular}
\end{table*}

We remark that the error in $C_{d,p}$ and $C_{d,v}$ present in the uncorrected values are determined primarily by the geometry of the roughness element.  Since we assume that the flow is viscous dominated in the interface region, the local flow conditions do not affect the partitioning of the total drag into viscous and pressure components. 

In order show the consistency of the presented formulations, we present results at different interface heights for elliptic roughness elements at $Re=40$ in table~\ref{tab:cyldragyi}. It can be clearly seen that there is no significant jump in errors of viscous and pressure drag at all the considered interface height. With increasing interface height,  there is a uniform increase in error.  Errors in velocity components show similar behaviour.  The same conclusions can be drawn by investigating data at different Re for square and elliptic roughness elements. 


The TR model~\cite{lacis2020} is derived based on the assumption that the microscopic Reynolds number $Re_s\le \eta^2$.  Microlength scale and the maximum slip velocity at the interface are used to define $Re_s$. For flow over the rough cylinder with elliptic inclusions,  $Re=40$ corresponds to $Re_s=0.2332$ which is approximately two orders of magnitude larger than $\eta^2$.  Even in this case, the velocity components along the interface and as a consequence drag components are predicted accurately, as shown in figure~\ref{fig:polarplot} and table~\ref{tab:cyldragres}. A similar conclusion is also reported by Sudhakar et al.~\cite{sudhakar2021} For flow over a cylinder at $Re=40$,  it is well known that the boundary layer separation occurs and a pair of attached vortices form. Such flow features are determined by $Re$, not by $Re_s$, and hence the effective model works well, as shown in this section.

\section{Conclusion}
\label{sec:conclusion}
This paper formulated the Transpiration Resistance (TR) model proposed for simulating flows over complex surfaces in the polar coordinate system. While existing effective models deal primarily with flat interfaces, we presented accurate predictions of the interface velocity over a circular interface using the proposed formulations. Moreover, we proposed constitutive parameters, called stress correction factors, which enable the accurate prediction of viscous and pressure drag components in the effective modelling framework. Also, we provided open-source codes used to simulate the flow over rough cylinders. While this paper used the TR model as the base formulation to show our results on force computation, any effective/homogenized model can be augmented with the stress correction factors proposed in this work for computing drag components on rough surfaces. 
\begin{acknowledgments}
We acknowledge the financial support provided by the DST-SERB Ramanujan fellowship~(sanction order no. SB/S2/RJN-037/2018).
\end{acknowledgments}

\section*{AUTHOR DECLARATIONS}
The authors have no conflicts to disclose.

\section*{Data Availability Statement}
Source codes to perform DNS, computation of constitutive parameters, and effective simulations of rough cylinders are made available in a public repository\cite{bitbucket}.
\appendix
\section{Effect of curvature in microscale problems for the circular interface}
In this paper, we proposed interface conditions in polar coordinates, and solved the microscale problems by taking the interface curvature into account~(figure~\ref{fig:cylmicroprob}).  Earlier studies\cite{zampogna2019,zampogna2020,ledda2021} reported simulations over circular interfaces, neglecting the effect of curvature in the solution of microscale problems. 
This appendix quantifies the effect of neglecting the curvature in solving the flow over a cylinder with elliptic roughness at $Re=40$. 

\begin{table*}
\caption{\label{tab:app_coef} Constitutive parameters for the circular cylinder problem with elliptic roughness.  The curvature is neglected in the computation of these parameters.}
\begin{ruledtabular}
\begin{tabular}{ccccc}
$h$ &\multicolumn{4}{c}{Constitutive coefficients}\\ \cline{2-5}
& $\mathcal{L}_{\theta\theta}$ & $\mathcal{M}$ & $\mathcal{P}_c$ & $\mathcal{S}_c$  \\ \hline
0        & 0.0888  &   0.0087  &  0.3682  &  0.6289 \\
0.05  & 0.1388  &   0.0092  &  0.3682  &  0.6317 \\
0.1     & 0.1888  &   0.0107  &  0.3682  &  0.6316 \\
0.2    & 0.2888  &   0.0143  &  0.3682  &  0.6316 \\
0.3   & 0.3888  &   0.0182  &  0.3682  &  0.6316 \\
\end{tabular}
\end{ruledtabular}
\end{table*}

In order to compute constitutive coefficients without  curvature effects, we consider the elliptic roughness element in polar coordinates~(figure~\ref{fig:cyldomain}(b)) and construct an equivalent microscale problem in Cartesian coordinates.  The dimensions of the microscale problem, illustrated in figure~\ref{fig:couetmicroprob}(b), are computed as follows: $l=r_b\alpha$, $r_x=r_b\beta$, and $r_y=d$.  The computed coefficients are presented in table~\ref{tab:app_coef}.  When compared to the values reported in table~\ref{tab:miccyl}, we can infer that neglecting the curvature leads to noticeable deviation in $\mathcal{L}_{\theta\theta}$ and $\mathcal{M}$ when the interface height $h\le 0.1$,  and negligible variation when the interface is moved away from the top surface of roughness. This observation is also reflected in errors in the interface velocity as presented in table~\ref{tab:app_err}. The error is computed by performing effective simulations using coefficients reported in table~\ref{tab:app_coef} and comparing the results with DNS.

Based on the presented data, we can conclude that curvature effects are important when the interface in the effective model is placed very close to the rough surface. 

\begin{table*}
\caption{Influence of interface height in computing interface velocities and drag coefficients for flow over a cylinder with elliptic roughness elements at $Re=40$.  Curvature effects are neglected in the microscale problem} 
\label{tab:app_err}
\begin{ruledtabular}
\begin{tabular}{ccccc}
$h$ &  \multicolumn{4}{c}{Error ($\%$)} \\ \cline{2-5}
& $u_\theta^{min}$ & $u_r^{max}$ & $C_{d,v}$ & $C_{d,p}$  \\ \hline
0         & 12.50   &  26.19   &  2.52    &  1.83 \\
0.05   & 10.39  &  22.30   &  3.18    &  1.12 \\
0.1      & 9.95    &  19.94   &  3.70    &  1.29 \\
0.2     & 10.58   &  18.14   &  4.66    &  1.65 \\
0.3     & 11.83   &  18.33  &  5.53    &  2.05 \\
\end{tabular}
\end{ruledtabular}
\end{table*}

\nocite{*}
\bibliography{cyl}
\end{document}